\documentclass[sigplan,10pt,screen,nonacm=true]{acmart}
\acmConference[PL'18]{ACM SIGPLAN Conference on Programming Languages}{January 01--03, 2018}{New York, NY, USA}
\acmYear{2018}
\acmISBN{} % \acmISBN{978-x-xxxx-xxxx-x/YY/MM}
\acmDOI{} % \acmDOI{10.1145/nnnnnnn.nnnnnnn}
\startPage{1}

\setcopyright{none}
\usepackage{booktabs}
\usepackage{graphicx}
\usepackage{amsmath}
\usepackage{algorithmic}
\usepackage{algorithm}
\usepackage{subcaption}
\usepackage{url}
\usepackage{siunitx}
\usepackage{xspace}
\usepackage{listings}
\usepackage{microtype}
\usepackage{multirow}
\usepackage{multicol}
\usepackage{tabularx}
\usepackage{mathtools}
\usepackage{enumitem}
\usepackage{adjustbox}
\usepackage{wrapfig}
\usepackage{syntax}
\usepackage{color}
\usepackage{colortbl}
\usepackage{placeins}
\usepackage{hyperref}

\newcommand{\reduce}[1]{\mathrel{#1}=}
\newcommand{\iteration}[1]{\forall_{#1} \;}
\newcommand{\titeration}[1]{\forall_{#1}}

\definecolor{todocolor}{rgb}{0.8,0,0}
\definecolor{keywordcolor}{rgb}{0.5,0,0.5}
\definecolor{nonsymmcolor}{gray}{0.9}

\newcommand{\figref}[1]{Figure~\ref{fig:#1}}
\newcommand{\figsref}[2]{Figures~\ref{fig:#1} and~\ref{fig:#2}}

\newcommand{\secref}[1]{Section~\ref{sec:#1}}

\newcommand{\tabref}[1]{Table~\ref{tab:#1}}
\newcommand{\HIDE}[1]{}

\ifdefined\notodo
\newcommand{\TODO}[1]{}
\newcommand{\saman}[1]{}
\else
\newcommand{\TODO}[1]{{\color{todocolor}#1}}
\newcommand{\saman}[1]{{\color{purple}Saman: #1}}
\fi

\newcolumntype{R}[2]{%
  >{\adjustbox{angle=#1,lap=\width-(#2)}\bgroup}%
  l%
  <{\egroup}%
}
\newcolumntype{R}{>{\raggedleft\arraybackslash}X}

\definecolor{textgray}{gray}{0.4}
\lstdefinestyle{nodedef}{
  mathescape,
  frame=none,
  aboveskip=\medskipamount,
  belowskip=\medskipamount,
  columns=flexible,
  basicstyle=\fontsize{8}{8}\ttfamily,
  keywords={def,supertype,elem,nonempty,seq,def,prototype,parent,size,in,bool},
  keywordstyle=\color{keywordcolor},
  commentstyle=\color{gray},
  keepspaces=true,
}
\newcommand\code[1]{\lstinline[columns=fullflexible, mathescape, basicstyle=\fontsize{9}{9}\ttfamily]|#1|}
\newcommand\smallcode[1]{\lstinline[columns=fullflexible, mathescape, basicstyle=\fontsize{8}{8}\ttfamily]|#1|}
\newcommand\smallnodedef[1]{\lstinline[columns=fullflexible, mathescape, keywords={def,supertype,elem,nonempty,seq,def,prototype,parent,size,in,bool}, basicstyle=\fontsize{8}{8}\ttfamily]|#1|}

\SetEnumitemKey{twocol}{
  before=\raggedcolumns\setlength{\multicolsep}{\topsep}\begin{multicols}{2},
  after=\end{multicols}
}

\begin{document}

\title{Dynamic Sparse Tensor Algebra Compilation}

\author{Stephen Chou}
\affiliation{
  \institution{MIT CSAIL}
  \streetaddress{32-G778, 32 Vassar Street}
  \city{Cambridge}
  \state{MA}
  \postcode{02139}
  \country{USA}
}
\email{s3chou@csail.mit.edu}

\author{Saman Amarasinghe}
\affiliation{
  \institution{MIT CSAIL}
  \streetaddress{32-G744, 32 Vassar Street}
  \city{Cambridge}
  \state{MA}
  \postcode{02139}
  \country{USA}
}
\email{saman@csail.mit.edu}

\begin{abstract}

This paper shows how to generate efficient tensor algebra code that compute on dynamic sparse tensors, which have sparsity structures that evolve over time.
We propose a language for precisely specifying recursive, pointer-based data structures, and we show how this language can express a wide range of dynamic data structures that support efficient modification, such as linked lists, binary search trees, and B-trees.
We then describe how, given high-level specifications of such data structures, a compiler can generate code to efficiently iterate over and compute with dynamic sparse tensors that are stored in the aforementioned data structures.
Furthermore, we define an abstract interface that captures how nonzeros can be inserted into dynamic data structures, and we show how this abstraction guides a compiler to emit efficient code that store the results of sparse tensor algebra computations in dynamic data structures.

We evaluate our technique and find that it generates efficient dynamic sparse tensor algebra kernels.
Code that our technique emits to compute the main kernel of the PageRank algorithm is 1.05$\times$ as fast as Aspen, a state-of-the-art dynamic graph processing framework.
Furthermore, our technique outperforms PAM, a parallel ordered (key-value) maps library, by 7.40$\times$ when used to implement element-wise addition of a dynamic sparse matrix to a static sparse matrix.
  
\end{abstract}

\maketitle

\section{Introduction}
\label{sec:introduction}
Sparse matrices and tensors (multidimensional arrays) are the de-facto data representation in many domains, including graph analytics~\cite{mattson2013,ligra}, machine learning~\cite{Rajbhandari2017,park2016faster}, and many others.
Countless formats for efficiently storing sparse tensors in memory have been proposed~\cite{smith2015tensor,Baskaran2012,hicoo,sell,csr5,stinger,aspen,terrace}, and many of these formats are supported by widely used sparse linear/tensor algebra libraries like Intel oneMKL~\cite{mkl} and graph processing frameworks like Ligra~\cite{ligra}.

\begin{figure}
  \centering
  \begin{minipage}[t]{0.49\columnwidth}
    \centering
    \includegraphics[scale=0.4]{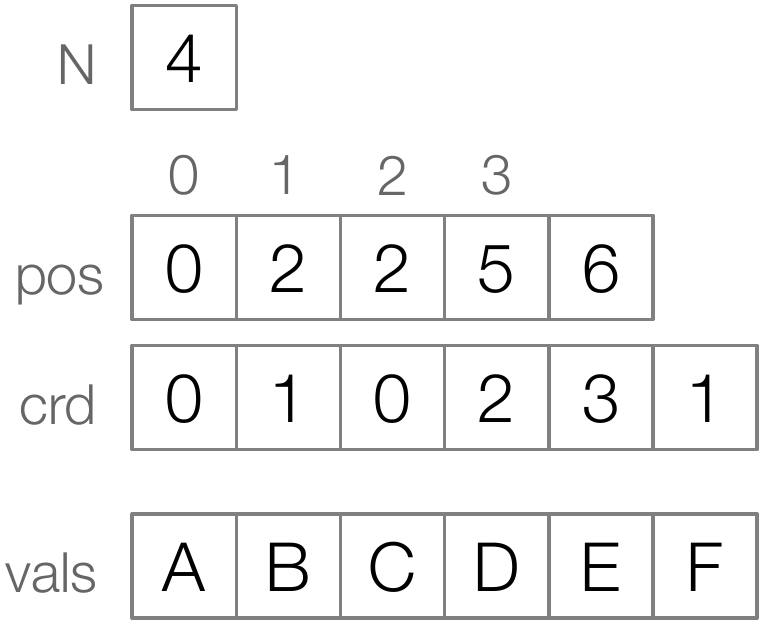}
    %\vspace{-12pt}
    \subcaption {
      \label{fig:csr-matrix-example}
      Sparse matrix stored in CSR
    }
    \vspace{8pt}
  \end{minipage}
  \hfill
  \begin{minipage}[t]{0.49\columnwidth}
    \centering
    \includegraphics[scale=0.4]{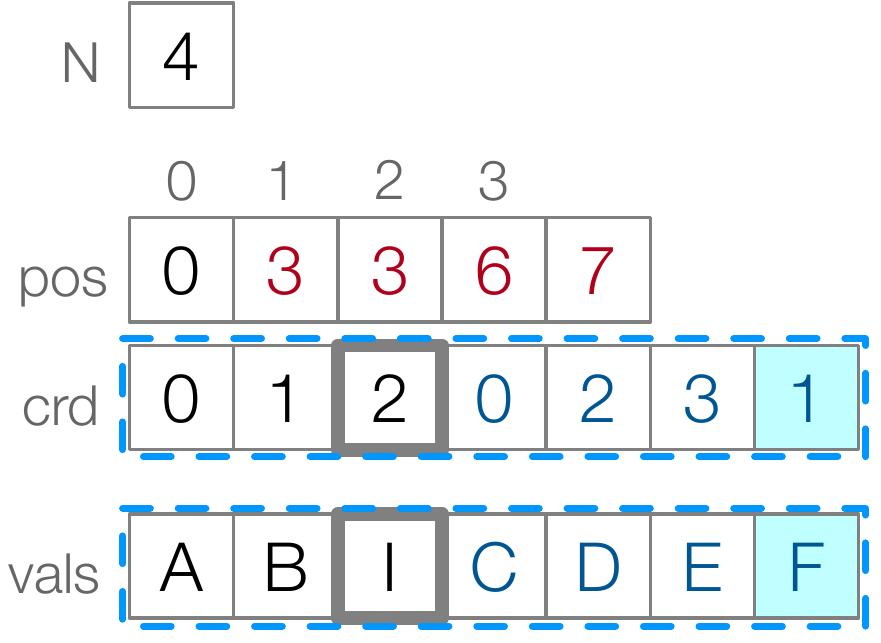}
    %\vspace{-12pt}
    \subcaption {
      \label{fig:csr-after-insert}
      CSR matrix after insertion
    }
    \vspace{8pt}
  \end{minipage}
  \begin{minipage}[t]{0.49\columnwidth}
    \centering
    \includegraphics[scale=0.4]{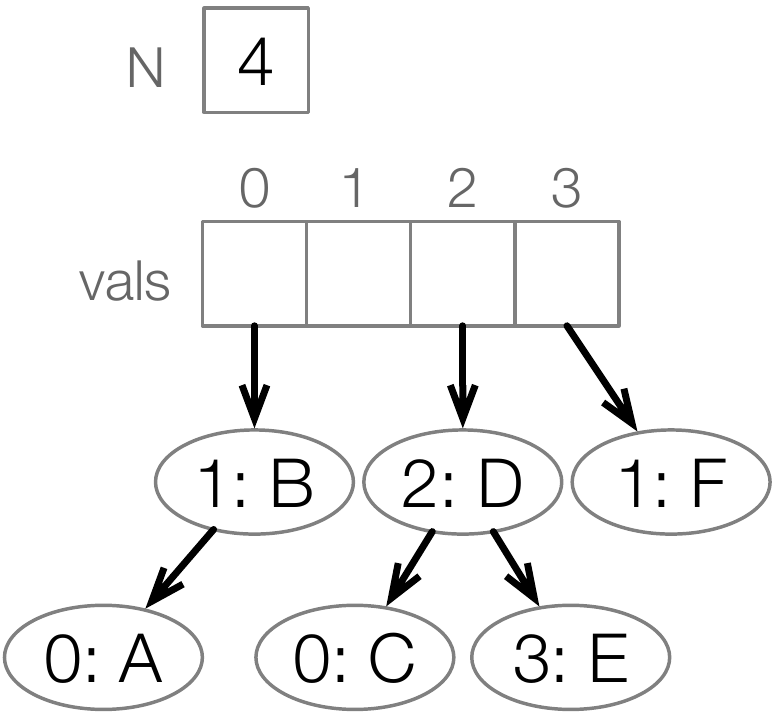}
    %\vspace{-12pt}
    \subcaption {
      \label{fig:bst-matrix-example}
      Sparse matrix stored as BSTs
    }
  \end{minipage}
  \hfill
  \begin{minipage}[t]{0.49\columnwidth}
    \centering
    %\vspace{-12pt}
    \includegraphics[scale=0.4]{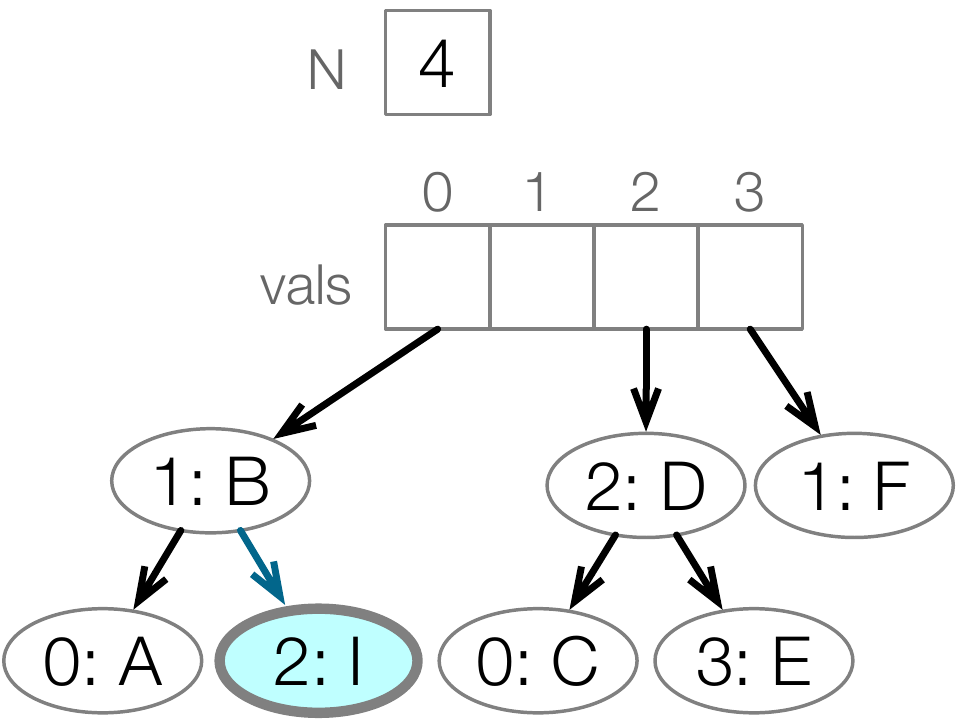}
    \subcaption {
      \label{fig:bst-after-insert}
      BST matrix after insertion
    }
  \end{minipage}
  \caption {
    Examples of the same matrix stored in CSR and as BSTs.
    Inserting a new nonzero {\bf I} into CSR at coordinates $(0, 2)$ requires shifting stored nonzeros in memory and may require reallocating the \code{crd} and \code{vals} arrays, whereas inserting the same nonzero into a BST only requires allocating a new node.
    }
  \label{fig:map-function-examples}
\end{figure}

Being able to efficiently compute with sparse tensors is crucial since real-world applications often work with large data sets, and lots of research effort has been devoted to optimizing the performance of sparse tensor computations~\cite{cagra,azad2017,hong2019,bell2008,smith2015splatt}.
High-performance libraries like oneMKL can efficiently compute with sparse tensors by storing them in formats like compressed sparse row (CSR) that use arrays to compactly store nonzeros in memory.
Such array-based formats provide good cache spatial locality and are ideal for storing \emph{static} sparse tensors that have constant sparsity structures (i.e., they do not gain new nonzeros over the course of an application's execution).
However, it is generally inefficient to insert a new nonzero into a tensor that is stored in a static sparse tensor format, since this may require already-stored nonzeros to be moved around in memory.
As \figref{csr-after-insert} illustrates, for instance, in order to insert a nonzero into a CSR matrix, all subsequent nonzeros must be shifted back so that space can be made for the new nonzero.

However, many real-world applications work with \emph{dynamic} sparse tensors that have constantly-evolving sparsity structures.
For instance, a sparse tensor that encodes friendship relations in a social network may regularly gain new nonzeros, reflecting new friendships that are formed between users over time.
Dynamic sparse tensor formats allow new nonzeros to be more efficiently inserted, since these formats use pointers to link together stored nonzeros and thus do not have to compact the nonzeros in memory.
\figref{bst-matrix-example}, for example, shows a dynamic tensor format that uses a binary search tree (BST) to store each row of a tensor.
Since nodes in a BST do not have to be stored contiguously in memory, a new nonzero can be inserted by simply allocating a new node and linking it to the rest of the BST without moving any existing node in memory.
Dynamic (pointer-based) tensor formats are typically less efficient to compute with than static (array-based) formats.
Nevertheless, since converting a tensor between formats can incur significant overhead~\cite{Xie2018}, an application that has to compute on dynamic sparse tensors can often do this more efficiently by just keeping the tensors stored in dynamic tensor formats.

\begin{figure}
  \centering
  \begin{minipage}[t]{0.48\columnwidth}
    \centering
    %\begin{tabular}{c}
    \begin{lstlisting}[basicstyle=\fontsize{8}{8}\ttfamily]
void map_b(blist* b,
    double* c, double& a) {
  while (b) {
    for (int32_t p = 0; 
         p < b->B; p++) {
      int32_t i = b->ec[p];
      a += b->ev[p] * c[i];
    }
    b = b->n;
  }
}

void compute(...) {
  for (int32_t i = 0; 
       i < N; i++) {
    double sum = 0.0;
    map_b(b[i]->h, c, sum);
    a[i] = sum;
  }
}
    \end{lstlisting}
    %\end{tabular}
    \vspace{-12pt}
    \subcaption {
      \label{fig:blist-spmv-example}
      SpMV with a sparse matrix stored as block linked lists.
    }
  \end{minipage}
  \hfill
  \begin{minipage}[t]{0.48\columnwidth}
    \centering
    %\begin{tabular}{c}
    \begin{lstlisting}[basicstyle=\fontsize{8}{8}\ttfamily]
void map_b(bst* b, 
    double* c, double& a) {
  if (b) {
    int32_t i = b->ec;
    a += b->ev * c[i];
    map_b(b->l, a, c);
    map_b(b->r, a, c);
  }
}

void compute(...) {
  for (int32_t i = 0; 
       i < N; i++) {
    double sum = 0.0;
    map_b(b[i]->r, c, sum);
    a[i] = sum;
  }
}


    \end{lstlisting}
    %\end{tabular}
    \vspace{-12pt}
    \subcaption {
      \label{fig:bst-spmv-example}
      SpMV with a sparse matrix stored as BSTs.
    }
  \end{minipage}
  \begin{minipage}[t]{\columnwidth}
    \centering
    %\begin{tabular}{c}
    \begin{minipage}[t]{0.48\linewidth}
    \centering
      \begin{lstlisting}[basicstyle=\fontsize{8}{8}\ttfamily]
      
inline uint8_t
iter_bst(uint8_t st, bst*& n,
    call_stack<...>& s,
    int32_t& c, double& v) {
  if (st == 1)
    goto iter_resume1;
  s.emplace(0, n);
  while (!s.empty()) {
    n = get<1>(s.top());
    if (get<0>(s.top()) == 1)
      goto call_resume1;
    while (n) {
      if (n->l) {
        get<0>(s.top()) = 1;
        get<1>(s.top()) = n;
        s.emplace(0, n->l);
        goto call_end;
call_resume1:;
      }
      c = n->ec; v = n->ev;
      return 1;
iter_resume1:
      n = n->r;
    }
    s.pop();
call_end:;
  }
  return 0;
}
      \end{lstlisting}
    \end{minipage}
    \hfill
    \begin{minipage}[t]{0.49\linewidth}
    \centering
      \begin{lstlisting}[basicstyle=\fontsize{8}{8}\ttfamily]

void compute(...) {
  ...
  for (int32_t i = 0; 
       i < N; i++) {
    bst* bn = b[i]->r;
    bst* cn = c[i]->r;
    uint8_t bstate = iter_bst(
      0, bn, bstack, jb, bval);
    uint8_t cstate = iter_bst(
      0, cn, cstack, jc, cval);
    while (bstate && cstate) {
      int32_t j = min(jb, jc);
      if (j == jb && j == jc)
        a[pa++] = bval * cval;
      if (j == jb)
        bstate = iter_bst(
          bstate, ..., bval);
      if (j == jc)
        cstate = iter_bst(
          cstate, ..., cval);
    }
  }
}
      \end{lstlisting}
    \end{minipage}
    %\end{tabular}
    \vspace{-12pt}
    \subcaption {
      \label{fig:bst-elwise-mul-example}
      Element-wise multiplication of sparse matrices stored as BSTs
    }
  \end{minipage}
  \caption {
    Examples of different dynamic sparse tensor algebra kernels with operands in disparate formats.
    Our technique is able to automatically generate all of these kernels.
  }
  \label{fig:compute-code-examples}
\end{figure}

There exist many distinct dynamic sparse tensor formats though, and they all have different trade-offs.
A format that uses BSTs to store nonzeros, for instance, can be efficiently modified but is also relatively inefficient to iterate over, since the CPU cache cannot be effectively utilized when accessing nonzeros.
Conversely, a format that uses blocked data structures like B-trees to store nonzeros can be more efficiently accessed, since some nonzeros are stored close together in memory (which increases cache utilization).
For the same reason, however, such a format cannot be as efficiently modified.
Thus, to be able to support a wide range of applications that have different proportions of data modification and compute, a sparse tensor algebra system must be able to efficiently compute with many disparate dynamic tensor formats.

Unfortunately, existing libraries that compute with dynamic sparse tensors (or graphs, which can also be modeled as tensors) each typically only support a few dynamic tensor formats.
As \figsref{blist-spmv-example}{bst-spmv-example} show, performing the same computation on dynamic sparse tensors that are stored in different formats requires distinct code that can be difficult to implement and optimize correctly.
Furthermore, as \figsref{bst-spmv-example}{bst-elwise-mul-example} show, performing different tensor algebra computations can also require very distinct code, even if they all work with dynamic sparse tensors that are stored in the same format.
It is thus impractical for library developers to manually implement all the code that would be needed to compute with a wide range of dynamic tensor formats, which motivates the need for a technique that can instead automatically generate such code.
However, existing sparse linear and tensor algebra compilers like TACO~\cite{kjolstad2017,kjolstad2019,chou2018,chou2020} cannot readily, if at all, generate code to efficiently compute on tensors stored in disparate dynamic tensor formats.

We propose the first technique that, given high-level specifications of a wide range of dynamic sparse tensor formats, can automatically generate sparse tensor algebra kernels that efficiently compute on tensors stored in the aforementioned formats.
In particular, we propose a language for precisely specifying how nonzeros can be stored in recursive, pointer-based data structures such as BSTs and linked lists, which compose to form many known dynamic tensor formats.
We show how a compiler can use these specifications to generate iterators and map functions for the aforementioned data structures, and we further show how the compiler can emit code that invokes these generated helper functions to efficiently compute on dynamic sparse tensors.
Additionally, we propose an abstract interface that captures how dynamic data structures can be efficiently assembled, and we show how a compiler can use implementations of this interface to generate tensor algebra kernels that store results in dynamic tensor formats.
In summary, our contributions include the
\begin{description}
  \item[node schema language,] which lets users precisely define a wide range of dynamic data structures that can be used to store dynamic sparse tensors (\secref{node-schema-language}); an
  \item[assembly abstraction] that captures how dynamic data structures can be efficiently constructed (\secref{assembly-interface}); and 
  \item[code generation techniques] that, guided by the above abstractions, emits efficient code to compute tensor algebra operations on dynamic sparse tensors (\secref{code-generation}).
\end{description}

We have implemented our technique as a prototype extension to the TACO sparse tensor algebra compiler.
Our evaluation shows that our technique generates efficient dynamic sparse tensor algebra kernels.
In particular, code that our technique emits to compute the PageRank algorithm's main kernel is 1.05$\times$ as fast as Aspen~\cite{aspen}, a state-of-the-art dynamic graph processing framework.
Additionally, our technique can emit efficient code to simultaneously compute with static and dynamic sparse tensors.
This lets our technique outperform PAM~\cite{pam}, a parallel ordered (key-value) maps library, by 7.40$\times$ when used to implement element-wise addition of a dynamic sparse matrix to a static matrix.

\section{Background}
\label{sec:background}
In this section, we give a brief overview of some of the many formats that have been proposed for storing dynamic sparse tensors.
Additionally, we briefly describe the sparse tensor algebra compilation techniques of \citet{kjolstad2019,kjolstad2017} and \citet{chou2018}, which generate efficient code that compute on static sparse tensors stored in array-based formats.

\subsection{Dynamic Sparse Tensor Formats}
\label{sec:dynamic-formats-background}

\begin{figure*}
 \begin{minipage}[b]{0.14\linewidth}
   \centering
   \includegraphics[scale=0.26]{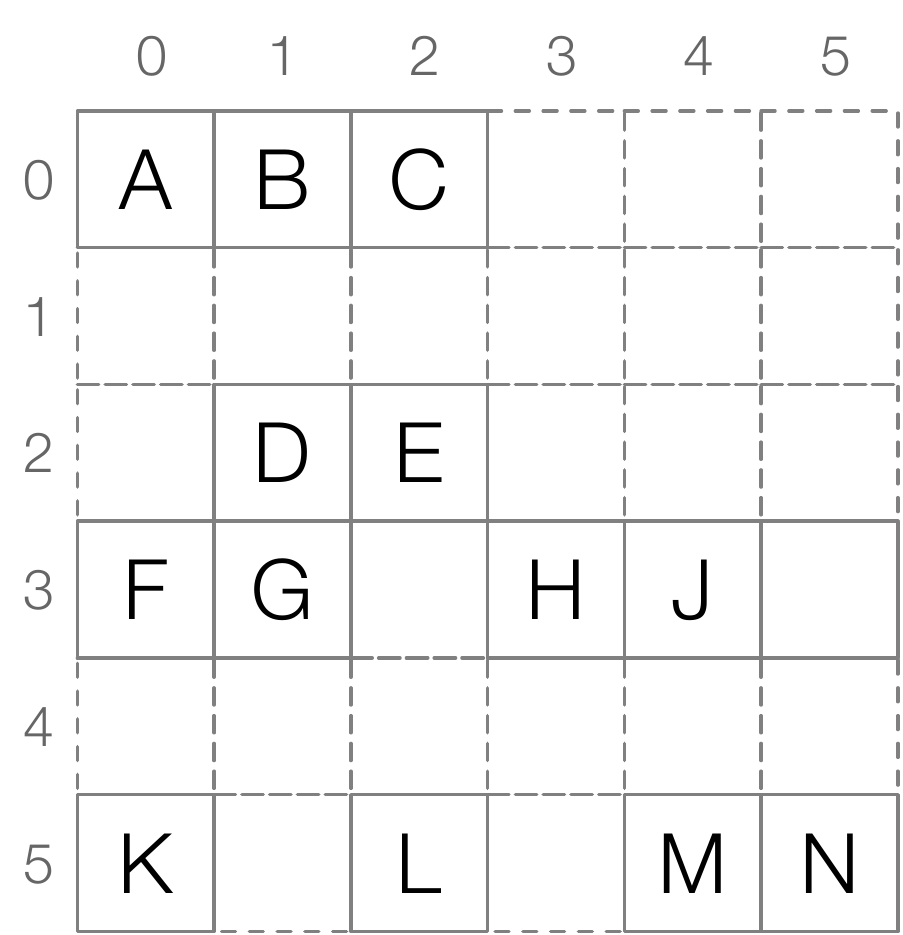}
   \subcaption{
     A 6$\times$6 tensor
   }
   \label{fig:matrix-example}
   \medskip
 \end{minipage}
 \hfill
 \begin{minipage}[b]{0.26\linewidth}
   \centering
   \includegraphics[scale=0.3]{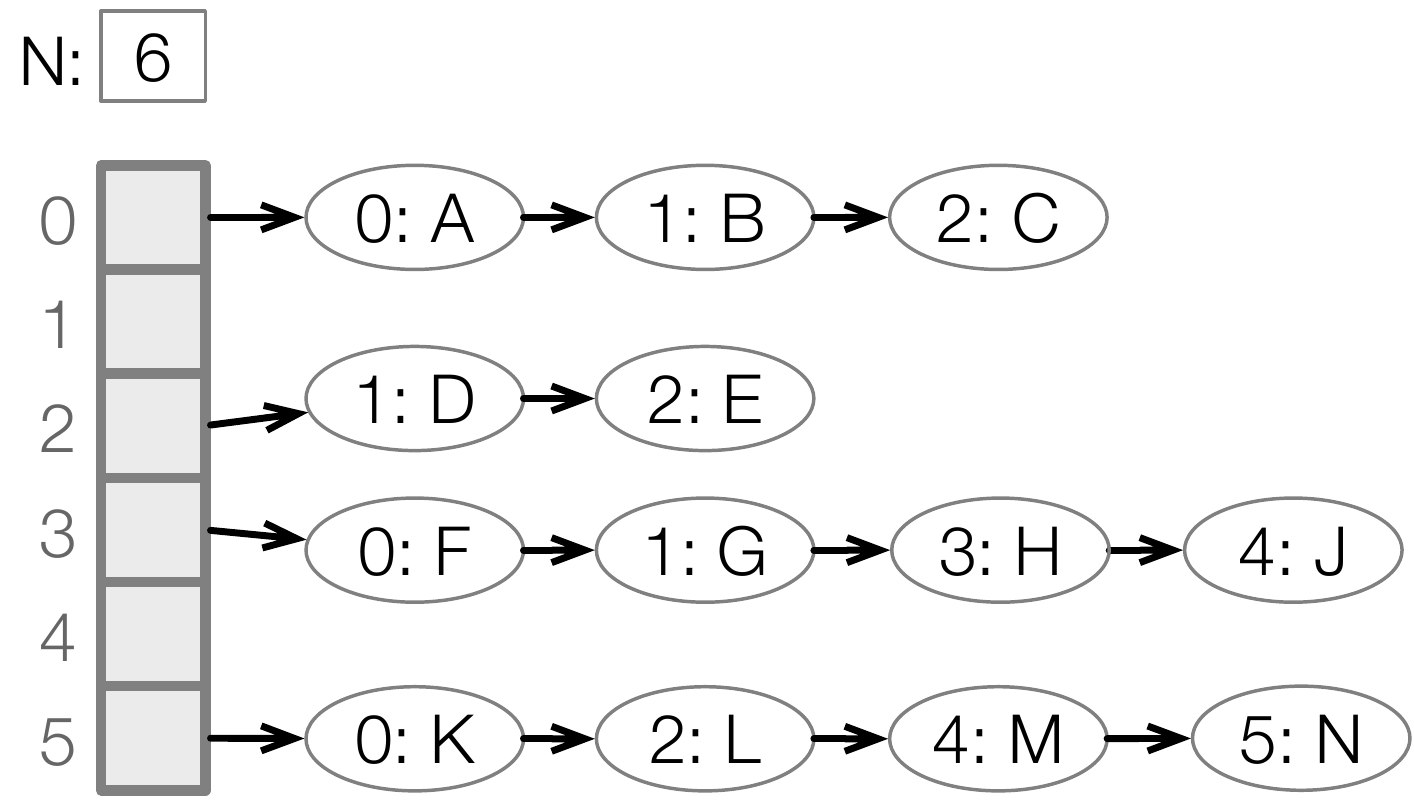}
   \subcaption{
     Adjacency list
   }
   \label{fig:matrix-example-adj-list}
   \medskip
 \end{minipage}
 \hfill
 \begin{minipage}[b]{0.28\linewidth}
   \centering
   \includegraphics[scale=0.3]{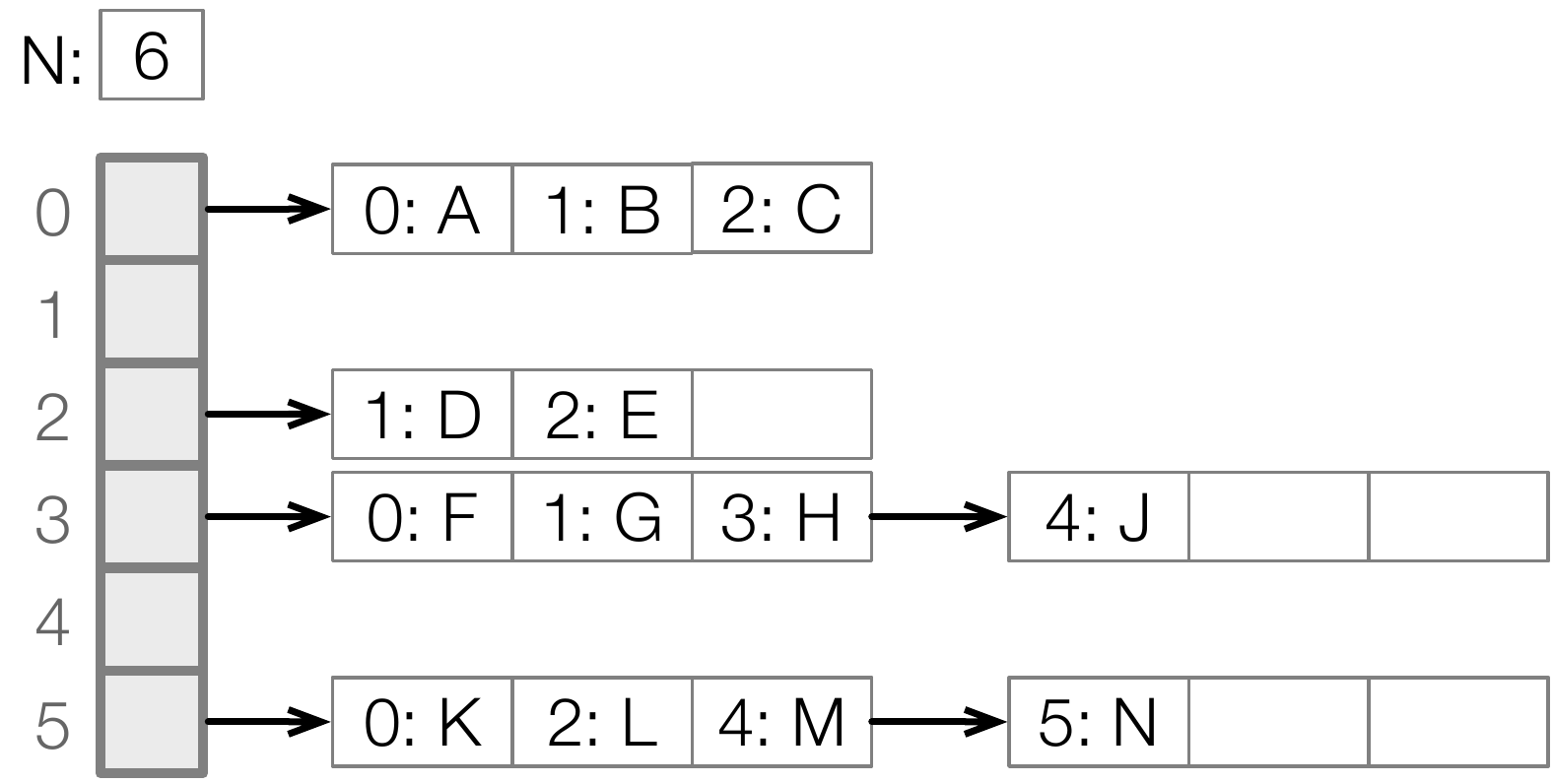}
   \subcaption{
     Block linked list
   }
   \label{fig:matrix-example-blist}
   \medskip
 \end{minipage}
 \hfill
 \begin{minipage}[b]{0.3\linewidth}
   \centering
   \includegraphics[scale=0.3]{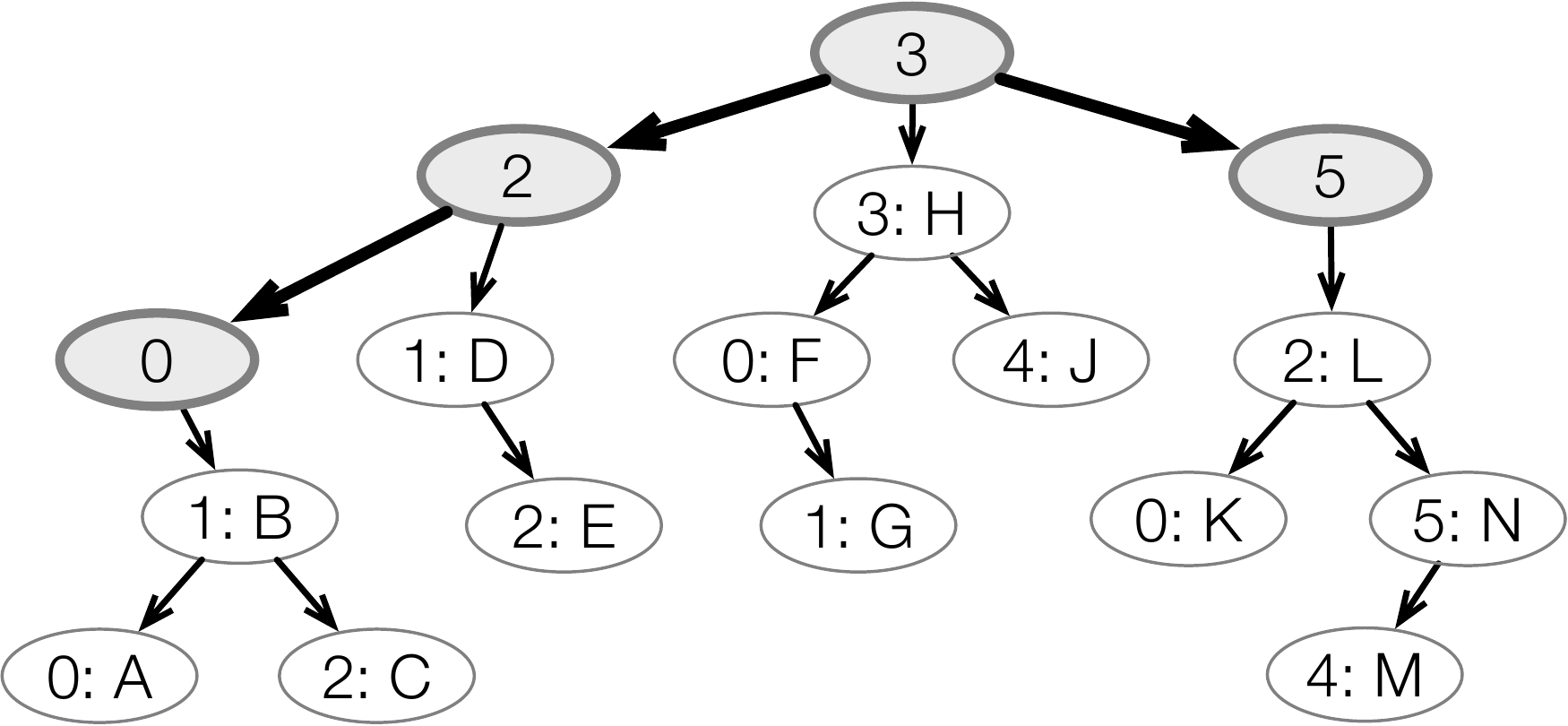}
   \subcaption{
    Binary search tree
   }
   \label{fig:matrix-example-bst}
   \medskip
 \end{minipage}
 \hfill
 \begin{minipage}[b]{0.24\linewidth}
   \centering
   \includegraphics[scale=0.3]{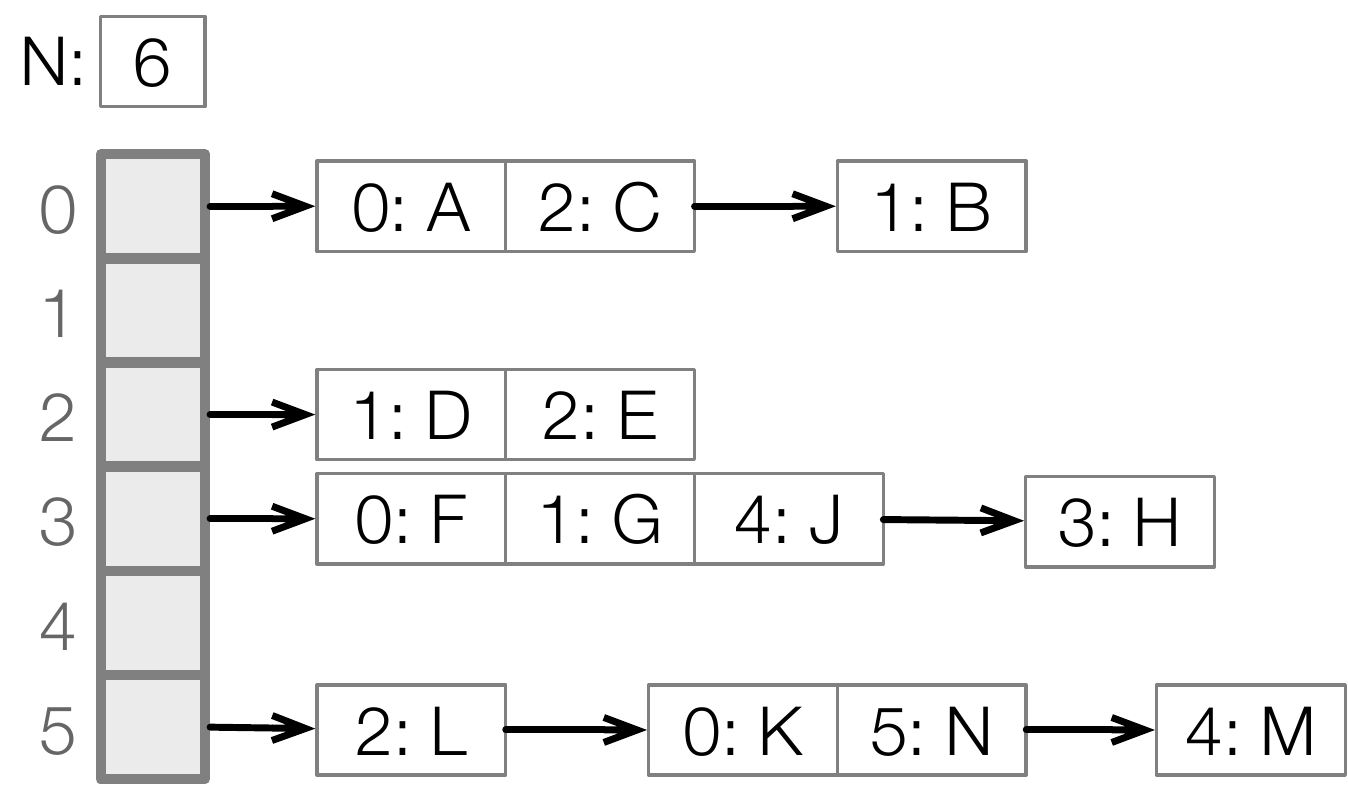}
   \subcaption{
     Variable block linked list
   }
   \label{fig:matrix-example-vblist}
   %\vspace*{-2mm}
 \end{minipage}
\hfill
 \begin{minipage}[b]{0.34\linewidth}
   \centering
   \includegraphics[scale=0.3]{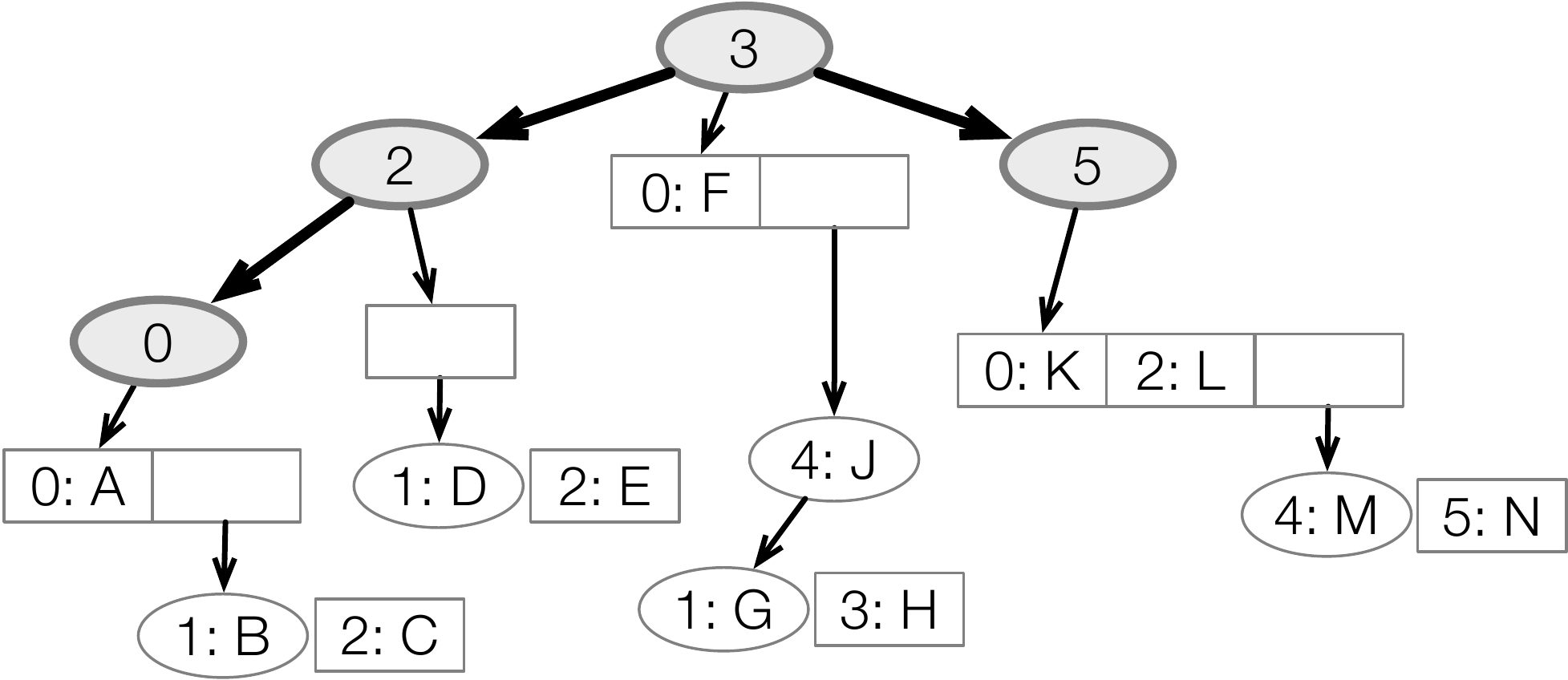}
   \subcaption{
     C-tree
   }
   \label{fig:matrix-example-ctree}
   %\vspace*{-2mm}
 \end{minipage}
 \hfill
 \begin{minipage}[b]{0.41\linewidth}
   \centering
   \medskip
   \includegraphics[scale=0.3]{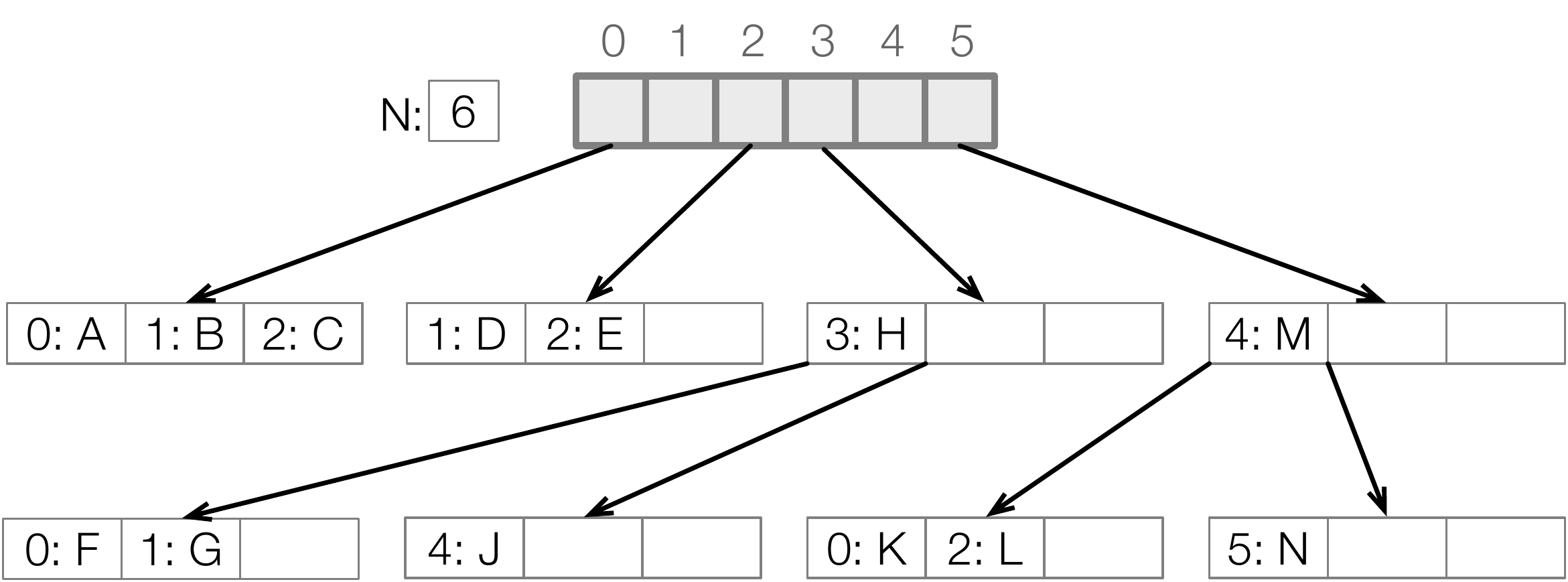}
   \subcaption{
     B-tree
   }
   \label{fig:matrix-example-btree}
   %\vspace*{-2mm}
 \end{minipage}
 \caption{
   Examples of disparate dynamic sparse tensor formats (b--g) storing the same tensor (a).
 }
 \label{fig:dynamic-tensor-formats}
\end{figure*}

There exists many formats for storing dynamic sparse tensors in memory, all of which possess different trade-offs.
\figref{dynamic-tensor-formats} shows several representative examples of these formats for two-dimensional dynamic tensors (i.e., matrices).

A standard way of representing any dynamic sparse matrix (such as adjacency matrices of dynamic graphs) is as a collection of adjacency lists, each of which stores the nonzeros in a single row of a matrix.
Each adjacency list can be stored as a linked list~\cite{clrs}, with each node in the linked list storing the column coordinate and value of one nonzero (\figref{matrix-example-adj-list}).
This representation enables new nonzeros to be efficiently added to a matrix by simply appending them to the appropriate adjacency lists, which can be done without moving any existing nonzero in memory.
Additionally, the collection of adjacency lists may itself be stored as a linked list, forming the list of lists representation; this enables new rows to be efficiently added to a matrix as well.

One drawback with linked lists though is that, when iterating over stored nonzeros, each access can incur a cache miss since nodes in a linked list may not be stored contiguously in memory.
This increases the overhead of accessing nonzeros and thus reduces performance when computing with dynamic tensors that are stored as linked lists.
To address this limitation, some high-performance graph processing frameworks like STINGER~\cite{stinger} instead use block linked lists that store multiple nonzeros in each node (\figref{matrix-example-blist}), effectively amortizing the overhead of each node access.
In a typical block linked list, every node contains an array of the same size and is able to store the same maximum number of nonzeros.
However, some frameworks like GraphOne~\cite{graphone} use variable block linked lists that allow different nodes to have arrays of different sizes, thus enabling nodes to store different maximum numbers of nonzeros (\figref{matrix-example-vblist}).
This allows updates to be efficiently batched, with every batch of new nonzeros inserted as just a single new node.

Another way of representing a dynamic sparse matrix is to use (balanced) binary search trees to store the set of nonzeros within each row as well as the set of nonempty rows (\figref{matrix-example-bst})~\cite{aspen}.
Using BSTs to store nonzeros enable new nonzeros to be efficiently inserted while also keeping the data structure sorted, which is useful for computations that require accessing nonzeros in order.
Again though, to amortize the overhead of accessing nodes in a tree, many high-performance graph processing frameworks instead use block tree data structures that store multiple nonzeros in each node.
For instance, Aspen~\cite{aspen} represents each row of a dynamic graph's adjacency matrix using a C-tree, which stores only a subset of nonzeros (or \emph{heads}) directly in a BST (\figref{matrix-example-ctree}).
The remaining nonzeros, meanwhile, are stored in either a \emph{prefix} (which contains all nonzeros that have smaller coordinates than any head element) or in chunks that are each associated with a distinct head element.
Similarly, Terrace~\cite{terrace} supports storing rows of an adjacency matrix using B-trees, which generalize BSTs in a different way by allowing each node to store more than two children in addition to also storing multiple nonzeros (\figref{matrix-example-btree}).

\subsection{Sparse Tensor Algebra Compilation}
\label{sec:tensor-compilation-background}

Our technique builds on the techniques of \citet{kjolstad2019,kjolstad2017} and \citet{chou2018}, which are implemented in the TACO sparse tensor algebra compiler.
TACO's code generator takes as input a tensor algebra computation expressed in \emph{concrete index notation}, which specifies how each entry in the output tensor should be computed in terms of entries in the input tensors.
(For example, matrix addition can be expressed in concrete index notation as $\titeration{i}\iteration{j} A_{ij} = B_{ij} + C_{ij}$, which specifies that each entry in the output tensor $A$ is the sum of the corresponding entries in the input tensors $B$ and $C$.)
Given such a concrete index notation statement, the code generator can recursively lower it to imperative code by emitting one or more loops to iterate over each dimension.
So to generate code that computes matrix addition, for instance, the code generator first emits one or more loops to iterate over all rows of $B$ and $C$ (i.e., dimension $i$).
Then, within each emitted loop over the rows of $B$ and $C$, the code generator emits one or more loops to iterate over all columns (i.e., dimension $j$) within a row in order to compute the element-wise sum of that row.

To generate code that compute with sparse tensors stored in specific formats though, TACO additionally requires the user to specify the format of each input and output tensor.
\citeauthor{chou2018} show how a wide range of static, array-based sparse tensor formats can be expressed as compositions of \emph{level formats}, each of which stores a dimension of a tensor.
The CSR format shown in \figref{csr-matrix-example}, for instance, can be expressed as a composition of two level formats \code{dense} and \code{compressed}, which store the row and column dimensions respectively.
The \code{dense} level format uses a single scalar variable \code{N} to encode a dense set of rows with coordinates from 0 to $N - 1$, while the \code{compressed} level format uses a \code{pos} array and a \code{crd} array to store the column coordinates of each row's nonzeros.
Both level formats---and all other level formats---implement snippets of imperative code that precisely describe how their underlying data structures can be accessed or assembled.
This lets TACO's code generator emit code to compute with tensors in specific formats by inlining the aforementioned code snippets into the generated loops.

\section{Dynamic Tensor Format Abstractions}
\label{sec:dynamic-format-abstractions}
In the same way that static tensor formats can be expressed as compositions of per-dimension level formats (as summarized in \secref{tensor-compilation-background}), dynamic tensor formats can also be expressed as compositions of per-dimension formats by generalizing level formats to support dynamic, pointer-based data structures.
Assume, for instance, we can define new level formats like \code{bst}, \code{ctree}, and \code{blist} that use BSTs, C-trees, and block linked lists to store a tensor dimension, respectively.
We can then express Aspen's adjacency matrix representation (\figref{matrix-example-ctree}) as (\code{bst}, \code{ctree}), indicating that the set of nonempty rows are stored using a BST while the set of nonzero columns for each row are stored using a C-tree.
Additionally, a tensor format may be composed of level formats that use static (array-based) and dynamic data structures.
For example, the composition (\code{dense}, \code{blist}) describes a tensor format that stores a matrix as a dense array of block linked lists, each of which stores a row of the matrix (\figref{matrix-example-blist}); this format is akin to STINGER's adjacency matrix representation.

In the rest of this section, we show how to precisely define level formats that store tensor dimensions using dynamic data structures.
In particular, we propose a new language called the \emph{node schema language}, and we show how a user can use this language to precisely specify how a dynamic data structure stores nonzeros (or nonempty subtensors) in memory (\secref{node-schema-language}).
We also show how, by implementing a common abstract interface that we define, a user can precisely specify how dynamic data structures are assembled (\secref{assembly-interface}).
As \secref{code-generation} will show, these specifications enable our technique to generate efficient code for computing on sparse tensors that are stored in dynamic tensor formats.

\subsection{Node Schema Language}
\label{sec:node-schema-language}

A wide range of dynamic tensor data structures, including all those described in \secref{dynamic-formats-background}, can be modeled as collections of nodes that are stored non-contiguously in memory, with each node storing a subset of nonzeros.
To precisely define any dynamic data structure, our technique requires a user to provide \emph{schemas} of the data structure's nodes, which specify how stored nonzeros are distributed amongst the nodes and how nodes are linked together.
These schemas can be expressed using the node schema language, the syntax for which is provided in \figref{node-schema-language-syntax}.

\begin{figure}
  {
    \small
    \setlength{\grammarparsep}{0.2em}
    \grammarindent1.08in
    \begin{grammar}
      <node_schema> ::= <supertype_def>$^*$ <node_def>$^+$
      
      <supertype_def> ::= `def' `supertype' <name>
      
      <node_def> ::= `def' <name> [`:' <name>] `{' <field_def>$^+$ [<sequence_def>] `}'
      
      <field_def> ::=  <name> `:' <type>
      
      <type> ::= <elem_type> | <child_type> | <size_type> \alt <metadata_type> | `parent'
      
      <elem_type> ::= `elem' [<array_type>] `nonempty'
      
      <child_type> ::= <name> [<array_type>] `nonempty'
      
      <array_type> ::= `[' (<name> | <const>) `]'
      
      <size_type> ::= `size' [`in' <array_size>]
      
      <array_size> ::= `[' <const> `,' (<const> | `*') `]'
      
      <metadata_type> ::= `bool' | `int8' | `uint8' | `int16' | `uint16' ...
      
      <sequence_def> ::= `seq' `=' <seq_entry> (`,' <seq_entry>)$^*$
      
      <seq_entry> ::= <name> | `{' <name> (`,' <name>)$^*$ `}'
    \end{grammar}
  }
  \caption {
    Syntax of the node schema language.
  }
  \label{fig:node-schema-language-syntax}
\end{figure}

\begin{figure}
  \centering
  \includegraphics[scale=0.4]{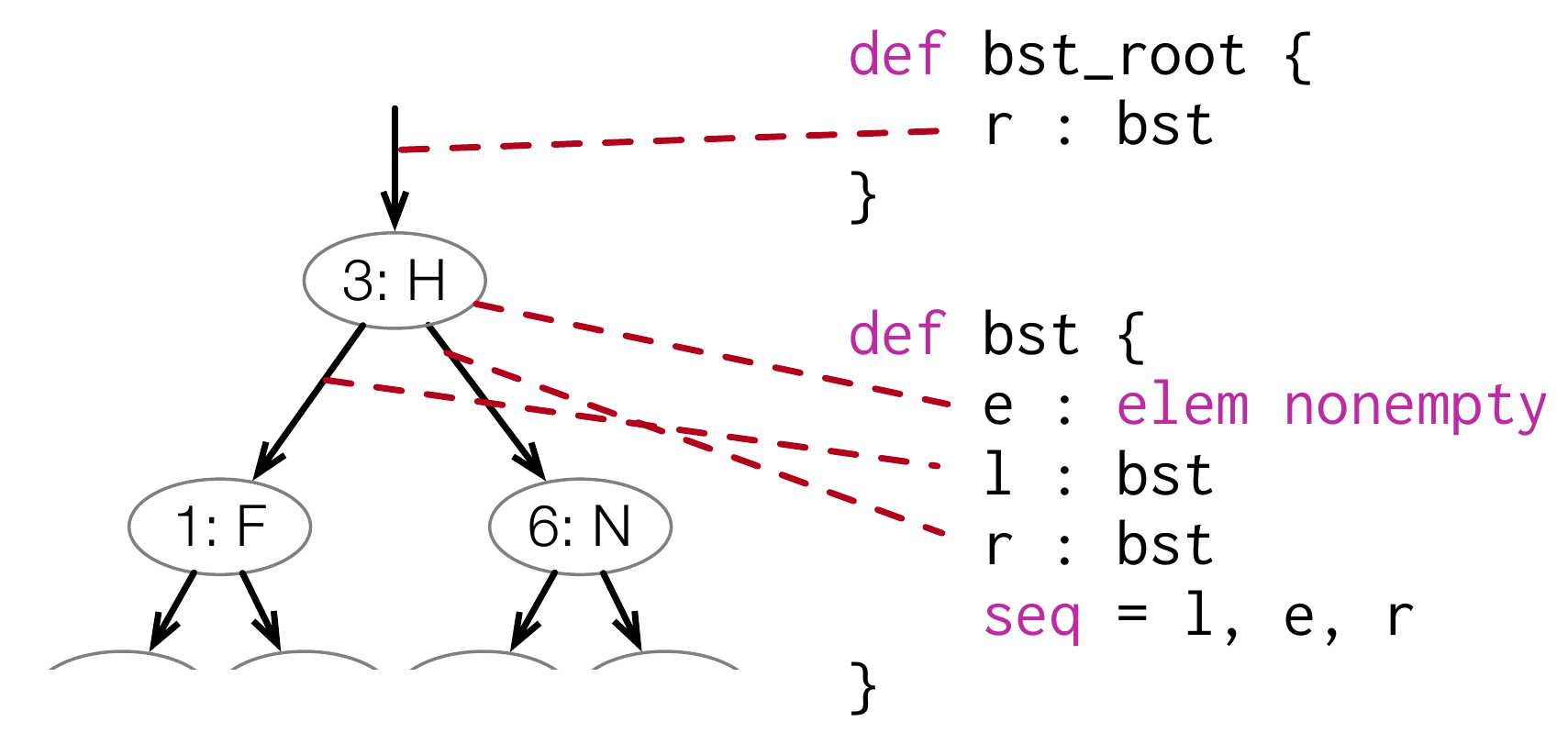}
  \caption {
    The node schemas for a BST precisely specifies how nonzeros are stored in nodes of a BST and how these nodes are linked together.
  }
  \label{fig:bst-schema}
\end{figure}

The node schema language allows users to define nodes that can contain an arbitrary number of fields, each of which may store nonzeros (or, more generally, nonempty subtensors) or store references to other nodes.
As an example, \figref{bst-schema} shows how binary search trees can be precisely defined using the node schema language.
In particular, a binary search tree consists of two types of nodes: a \code{bst_root} node, which simply stores a reference to the root of the tree, and \code{bst} nodes, which actually contain the nonzeros.
The schema for \code{bst} nodes specifies that each node stores one nonzero \code{e} as well as stores references to up to two child nodes \code{l} and \code{r}, both of which are of the same type.
(The \code{nonempty} annotation specifies that each node must store exactly one nonzero and cannot be empty.)
Furthermore, the schema contains a \emph{sequence attribute} (\code{seq}) that specifies the ordering of nonzeros stored by all reachable nodes; in particular, all nonzeros reachable from \code{l} must have coordinates less than that of \code{e}, which in turns must have coordinate less than those of all nonzeros reachable from \code{r}.
Meanwhile, the schema for the \code{bst_root} node simply specifies that it stores a reference to the root node \code{r}, which may be null if the tree is empty.

Nodes in a dynamic data structure may be defined to store more than one nonzero.
\figref{ttree-spec}, for instance, shows how T-trees~\cite{ttree}, which generalize BSTs by having each node store a bounded-size block of nonzeros, can be precisely defined.
In particular, the schema for \code{ttree} nodes specifies that each node can store multiple nonzeros contiguously in an array \code{e}, with the exact number of nonzeros that \code{e} contains being stored in a separate field \code{B}.
Different nodes may store different numbers of nonzeros, but the \code{in} clause (in the declaration of the \code{B} field) constrains each node to contain at least one and at most four nonzeros.
Like with BSTs, the sequence attribute specifies that all nonzeros stored in a node (in array \code{e}) have coordinates greater than those of all nonzeros stored in the left subtree \code{l} but less than those of all nonzeros stored in the right subtree \code{r}.
Additionally though, the \code{\{e\}} term in the sequence attribute indicates that nonzeros are stored within \code{e} in increasing order by their coordinates, so that \code{e[0]} stores the nonzero with the smallest coordinate, \code{e[1]} stores the nonzero with the second-smallest coordinate, and so on.
(In general, sequence attribute terms that are enclosed within braces may contain multiple arrays, which specifies the array elements are ordered in interleaved order.
So, for instance, the term \code{\{c, e\}} in the sequence attribute for internal B-tree nodes (\figref{btree-spec}) denotes that all nonzeros stored in the subtree \code{c[0]} have smaller coordinates than \code{e[0]}, which in turn has a smaller coordinate than all nonzeros in the subtree \code{c[1]}, and so on.)

\begin{figure}
  \centering
  \begin{minipage}[t]{0.48\columnwidth}
    \centering
    \begin{lstlisting}[style=nodedef]
def list {
  e : elem nonempty
  n : list
  seq = {e}, n
}

def list_head {
  h : list
}

    \end{lstlisting}
    \vspace{-12pt}
    \subcaption {
      \label{fig:list-spec}
      Linked list
    }
    \vspace{8pt}
  \end{minipage}
  \hfill
  \begin{minipage}[t]{0.48\columnwidth}
    \centering
    \begin{lstlisting}[style=nodedef]
def blist {
  e : elem[B] nonempty
  n : blist
  B : size in [0, 3]
  seq = {e}, n
}

def blist_head {
  h : blist
}
    \end{lstlisting}
    \vspace{-12pt}
    \subcaption {
      \label{fig:blist-spec}
      Block linked list
    }
     \vspace{8pt}
  \end{minipage}
  \hfill
  \begin{minipage}[t]{0.48\columnwidth}
    \centering
    \begin{lstlisting}[style=nodedef]
def vblist {
  e : elem[B] nonempty
  n : vblist
  B : size
  seq = {e}, n
}

def vblist_head {
  h : vblist
}

    \end{lstlisting}
    \vspace{-12pt}
    \subcaption {
      \label{fig:vblist-spec}
      Variable block linked list
    }
     \vspace{8pt}
  \end{minipage}
  \hfill
  \begin{minipage}[t]{0.48\columnwidth}
    \centering
    \begin{lstlisting}[style=nodedef]
def ttree {
  v : elem[B] nonempty
  l : ttree
  r : ttree
  B : size in [1, 4]
  seq = l, {e}, r
}

def ttree_root {
  r : ttree
}
    \end{lstlisting}
    \vspace{-12pt}
    \subcaption {
      \label{fig:ttree-spec}
      T-tree
    }
     \vspace{8pt}
  \end{minipage}
  \hfill
  \begin{minipage}[t]{0.48\columnwidth}
    \centering
    \begin{lstlisting}[style=nodedef]
def rbtree {
  e : elem nonempty
  l : rbtree
  r : rbtree
  p : parent
  c : bool
  seq = l, e, r
}

def rbtree_root {
  r : rbtree
}



    \end{lstlisting}
    \vspace{-12pt}
    \subcaption {
      \label{fig:rbtree-spec}
      Red-black tree
    }
     \vspace{8pt}
  \end{minipage}
  \hfill
  \begin{minipage}[t]{0.48\columnwidth}
    \centering
    \begin{lstlisting}[style=nodedef]
def ctree {
  h : elem nonempty
  t : elem[N] nonempty
  l : ctree
  r : ctree
  N : size
  seq = l, h, {t}, r
}

def prefix {
  e : elem[N] nonempty
  r : ctree
  N : size
  seq = {e}, r
}
    \end{lstlisting}
    \vspace{-12pt}
    \subcaption {
      \label{fig:ctree-spec}
      C-tree
    }
     \vspace{8pt}
  \end{minipage}
  \hfill
  \begin{minipage}[t]{\columnwidth}
    \centering
    \begin{lstlisting}[style=nodedef,multicols=2]
def supertype btree

def btree_internal : btree {
  e  : elem[B] nonempty
  c  : btree[B] nonempty
  cl : btree nonempty
  B  : size in [1, 3]
  seq = {c, e}, cl
}

def btree_leaf : btree {
  e : elem[B] nonempty
  B : size in [1, 3]
  seq = {e}
}

def btree_root {
  r : btree
}
    \end{lstlisting}
    \vspace{-6pt}
    \subcaption {
      \label{fig:btree-spec}
      B-tree
    }
  \end{minipage}
  \caption {
    Node schemas for a wide range of dynamic data structures, including all those in \figref{dynamic-tensor-formats}.
  }
  \label{fig:node-schema-examples}
\end{figure}

Annotations to node schemas and their fields, such as sequence attributes and \code{nonempty} annotations, are optional, which makes it possible to define many practical variants of a dynamic data structure.
\figsref{blist-spec}{blist-variants}, for instance, show how the node schema language can be used to define four variants of block linked lists, each of which pads blocks and orders stored nonzeros in a different way.
Similarly, \figref{ctree-spec} shows how a declaration of a \code{size} field can omit the \code{in} clause, indicating that the size of an array field is unconstrained.
This makes it possible to precisely define a C-tree, which is also a block BST data structure but, unlike T-trees, do not strictly limit the number of nonzeros stored in each node.

\begin{figure}
  \centering
  \begin{minipage}[t]{0.33\columnwidth}
    \centering
    %\begin{tabular}{c}
    \begin{lstlisting}[style=nodedef]
def blist {
  e : elem[B]
  n : blist
  B : size in [0, 3]
  seq = {e}, n
}
\end{lstlisting}
    \includegraphics[scale=0.36]{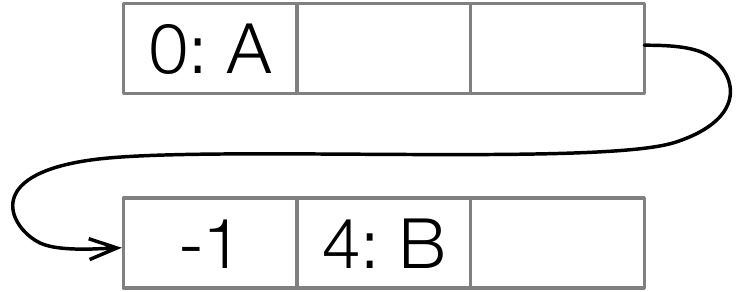}
    %\end{tabular}
    \subcaption {
      \label{fig:blist-nullable-partial}
      With empty slots up to position \smallcode{B}
    }
  \end{minipage}
  \hfill
  \begin{minipage}[t]{0.32\columnwidth}
    \centering
    %\begin{tabular}{c}
\begin{lstlisting}[style=nodedef,showlines=true]
def blist {
  e : elem[3]
  n : blist
  seq = {e}, n
}

\end{lstlisting}
    \includegraphics[scale=0.36]{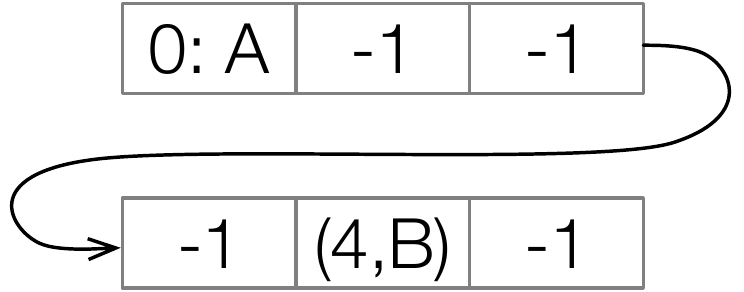}
    %\end{tabular}
    \subcaption {
      \label{fig:blist-nullable-full}
      With empty slots anywhere in block
    }
  \end{minipage}
    \hfill
\begin{minipage}[t]{0.31\columnwidth}
    \centering
    %\begin{tabular}{c}
\begin{lstlisting}[style=nodedef,showlines=true]
def blist {
  e : elem[3]
  n : blist
}


\end{lstlisting}
    \includegraphics[scale=0.36]{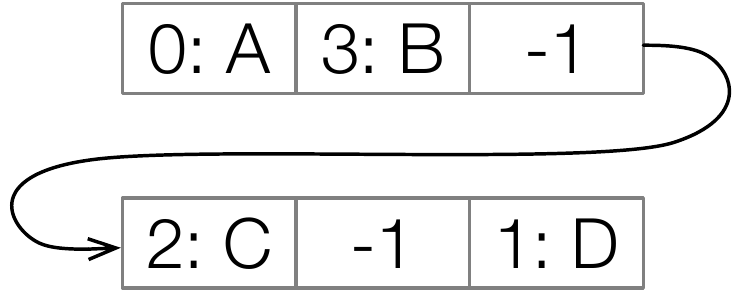}
    %\end{tabular}
    \subcaption {
      \label{fig:blist-nullable-unsorted}
      With nonzeros unsorted
    }
  \end{minipage}
  \caption {
    The node schema language can specify many variants of block linked lists. ("-1" denote slots with no nonzero.)
  }
  \label{fig:blist-variants}
\end{figure}

A dynamic data structure may consist of multiple types of nodes, all of which can potentially store nonzeros.
For instance, as \figref{matrix-example-ctree} illustrates, a C-tree typically has an associated prefix, which contains all nonzeros that precede the first head element (when ordered by their coordinates) and are thus stored separately from the actual tree data structure.
\figref{ctree-spec} shows, a C-tree with its associated prefix can be expressed in the node schema language by defining an additional \code{prefix} node that stores the prefix in an array \code{e} and that also stores a reference \code{r} to actual tree, which is represented by a node of a different type (i.e., \code{ctree}).

While different types of nodes may possess different sets of fields, they can nevertheless share a common supertype, which allows a single reference to point to a node that is of one of several different types.
For instance, B-trees consist of two types of nodes: internal nodes, which need to store references to child nodes, and child nodes, which can omit those references to reduce space usage.
As \figref{btree-spec} shows, by defining internal nodes (\code{btree_internal}) and leaf nodes (\code{btree_leaf}) to be of the same supertype \code{btree}, a user can specify that each child of an internal node may itself be another internal node or, alternatively, be a leaf node.

Finally, the node schema language allows users to specify that nodes store additional metadata, which may not be strictly needed to store nonzeros but are useful for other purposes.
For instance, \figref{rbtree-spec} shows how a node in a red-black tree can be defined to store a reference to its parent (in field \code{p}) as well as another field \code{c} that represents the node's color; these fields are needed to support efficient insertions into a red-black tree while keeping the tree balanced.

\subsection{Assembly Abstract Interface}
\label{sec:assembly-interface}

To support generating sparse tensor algebra kernels that store results in dynamic tensor formats, our technique additionally requires users to implement an abstraction that captures how dynamic data structures are efficiently assembled.
Specifically, for any dynamic data structure, a user must specify how nonzeros can be individually \emph{appended} to the data structure and/or specify how the data structure can be \emph{bulk assembled} from a set of nonzeros.

Appends to a dynamic data structure are defined by two functions in our abstraction:
\begin{itemize}[leftmargin=1.25\parindent]
  \item \code{append_first(elem, st, ret);}
  \item \code{append_rest(elem, st);}
\end{itemize}
\code{append_first} defines how the first nonzero can be appended to the data structure, while \code{append_rest} defines how all subsequent nonzeros can be appended (in order of their coordinates, if the data structure is specified to be sorted by a sequence attribute).
Both functions take as inputs the nonzero to be appended (\code{elem}) as well as a reference to a user-defined object (\code{st}) that can be used to keep track of where exactly a nonzero was last appended in the data structure being assembled.
Additionally, \code{append_first} takes as input a reference to a preallocated node (\code{ret}) that is intended to serve as a handle to the data structure to be assembled.
\figref{append-example} demonstrates how the append functions can be implemented for one specific dynamic data structure, namely block linked lists.
To append the first new nonzero, \code{append_first} for block linked lists allocates a block, stores the nonzero at the beginning of the block, and initializes the root pointer (\code{ret->h}) to point to the block.
For each subsequent new nonzero, \code{append_rest} then simply appends the new nonzero to the end of the last allocated block (which is cached in \code{st}) unless the block is already full, in which case a new block is first allocated and attached to the rest of the list.

\begin{figure}
  \centering
  \begin{minipage}[t]{0.48\columnwidth}
    \centering
    %\begin{tabular}{c}
    \begin{lstlisting}[basicstyle=\fontsize{8}{8}\ttfamily]
st = {
  node : blist
};

append_first(elem, st, ret):
  blist* node = new blist;
  node->e[0] = elem;
  node->B = 1;
  node->n = null;
  ret->h = node;
  st->node = node;
  
append_rest(elem, st):
  blist* node = st->node;
  if (node->B == 4) {
    node = new blist;
    node->B = 0;
    node->n = null;
    st->node->n = node;
    st->node = node;
  }
  node->e[node->B] = elem;
  node->B += 1; 
  
  
  
  
  
    \end{lstlisting}
    %\end{tabular}
      \vspace{-12pt}
    \subcaption {
      \label{fig:append-example}
      Append for block linked lists
    }
  \end{minipage}
  \hfill
  \begin{minipage}[t]{0.50\columnwidth}
    \centering
    %\begin{tabular}{c}
    \begin{lstlisting}[basicstyle=\fontsize{8}{8}\ttfamily]
build_rbt(elems, s, e):
  if (s > e)
    return null;
  rbtree* node = new rbtree;
  uint64 m = (s + e) / 2;
  node->v = elems[m];
  node->p = null;
  node->c = (s + 1 == e);
  if (s == e) {
    node->l = node->r = null;
  } else if (s + 1 == e) {
    node->l = 
      build_rbt(elems, s, s);
    node->r = null;
    node->l->p = node;
  } else {
    node->l = 
      build_rbt(elems, s, m-1);
    node->r =
      build_rbt(elems, m+1, e);
    node->l->p = node;
    node->r->p = node;
  }
  return node;

build(ret, elems, sz):
  ret->r =
    build_rbt(elems, 0, sz-1);
    \end{lstlisting}
    %\end{tabular}
    \vspace{-12pt}
    \subcaption {
      \label{fig:build-example}
      Bulk assembly for BSTs
    }
  \end{minipage}
  \caption {
    Examples of how various dynamic data structures can implement the assembly functions in our abstraction.
  }
  \label{fig:assembly-function-examples}
\end{figure}

Bulk assembly of a dynamic data structure, on the other hand, is defined by a single function in our abstraction
\begin{itemize}[leftmargin=1.25\parindent]
  \item \code{build(elems, sz, ret);}
\end{itemize}
where \code{elems} represents the sequence of nonzeros to be inserted, \code{sz} stores the size of \code{elems}, and \code{ret} is again a reference to a preallocated node that is intended to serve as a handle to the data structure being assembled.
\code{elems} implements an array interface, so any nonzero can be accessed by their position in the sequence.
Additionally, if the data structure being assembled is specified to be sorted (i.e., if stored nonzeros are ordered by a sequence attribute), then the nonzeros in \code{elems} are guaranteed to be ordered by their coordinates.
\figref{build-example} shows how a user can implement the \code{build} function for red-black trees.
Bulk assembly can often be implemented more efficiently than appends.
In the case of red-black trees, for instance, bulk assembly can be performed without needing to rebalance the tree for each inserted nonzero, which by contrast is needed when appending to red-black trees.
Furthermore, bulk assembly is typically more amenable to parallelization; for example, a user can trivially parallelize the implementation of \code{build} in \figref{build-example} by having recursive calls to \code{build_rbt} be spawned in parallel.
However, bulk assembly requires the set of inserted nonzeros (\code{elems}) to be fully precomputed, which for certain tensor algebra computations may incur additional overhead.

\section{Code Generation}
\label{sec:code-generation}
In this section, we describe how we generalize the techniques of \citet{kjolstad2019,kjolstad2017} and \citet{chou2018} to generate efficient code that compute on tensors stored in arbitrary combinations of dynamic and static tensor formats.
Like the technique of \citeauthor{kjolstad2019}, which was summarized in \secref{tensor-compilation-background}, our technique takes as input a tensor algebra computation expressed in concrete index notation and recursively emits imperative (C++) code to iterate over each dimension of the inputs. %and perform the computation.
The remainder of this section will thus focus on how our technique emits code to efficiently compute on nonzeros along just one dimension.
In particular, we show how our technique can use the abstractions we propose in \secref{dynamic-format-abstractions} to generate code that may be optimized in very different ways for different computations and dynamic tensor formats.
The result is a system that reduces the effort needed to efficiently compute with dynamic sparse tensors.

\subsection{Generating Node Declarations}
\label{sec:generating-nodes}

Before generating code to compute on dynamic sparse tensors, our technique first emits code to declare structs that represent nodes of dynamic data structures and that the generated code can actually work with.
These structs are directly generated from node schemas, with one struct generated for each node schema.
\figref{struct-examples} shows examples of structs that our technique generates for some of the dynamic data structures defined in \secref{node-schema-language}.

\begin{figure}
  \centering
  \begin{minipage}[t]{0.48\columnwidth}
    \centering
    %\begin{tabular}{c}
    \begin{minipage}[t]{\linewidth}
      \centering
      \begin{lstlisting}[basicstyle=\fontsize{8}{8}\ttfamily]
struct bst {
  int32_t ec;
  double ev;
  bst* r;
  bst* l;
};
      \end{lstlisting}
      \vspace{-12pt}
      \subcaption {
        \label{fig:bst-struct}
        BST
      }
      \vspace{14pt}
    \end{minipage}
    \begin{minipage}[t]{\linewidth}
      \centering
      \begin{lstlisting}[basicstyle=\fontsize{8}{8}\ttfamily]
struct blist {
  int32_t ec[4];
  double ev[4];
  blist* n;
  int32_t B;
};
      \end{lstlisting}
        \vspace{-12pt}
      \subcaption {
        \label{fig:blist-struct}
        Block linked list
      }
    \end{minipage}
    %\end{tabular}
  \end{minipage}
  \hfill
  \begin{minipage}[t]{0.48\columnwidth}
    \centering
    %\begin{tabular}{c}
    \begin{lstlisting}[basicstyle=\fontsize{8}{8}\ttfamily]
struct btree {
  enum type {
    btree_internal, 
    btree_leaf
  };
  type tp;
};

struct btree_internal 
    : public btree {
  int32_t ec[4];
  double ev[4];
  btree* c[4];
  btree* cl;
  int32_t B;
};
    \end{lstlisting}
    %\end{tabular}
    \vspace{-12pt}
    \subcaption {
      \label{fig:btree-struct}
      B-tree
    }
  \end{minipage}
  \caption {
    Examples of structs that our technique emits for storing various dynamic data structures.
  }
  \label{fig:struct-examples}
\end{figure}

\tabref{node-struct-fields} shows how our technique translates scalar fields in a node schema to fields in the corresponding struct.
Array fields are translated in the same way, except each emitted field is either an array member (e.g., \code{int32_t fc[4]}) or a pointer to an array (e.g., \code{int32_t* fc}).
In particular, if an array field's size is either a constant \code{N} or upper-bounded by a constant \code{N} (with an \code{in} clause), then the field is translated to an array member of size \code{N}; otherwise, the array field is translated to a pointer to an array.
Furthermore, to support nodes that share a common supertype, our technique emits a struct for each supertype \code{T}, and all of \code{T}'s subtypes have structs that inherit from \code{T}'s struct.
This struct contains a single member \code{tp}, which stores an enumeration that is intended for keeping track of a node's concrete type.

{
\setlength{\tabcolsep}{3.5pt}
\begin{table}
  \caption{
    Translation of (scalar) fields in a node schema to fields in the corresponding emitted struct.
  }
  \centering
  {\small
  \begin{tabularx}{\columnwidth}{llX}
    \toprule
    \multicolumn{1}{c}{Field in Schema} & \multicolumn{1}{c}{\begin{tabular}{@{}c@{}}Emitted Field(s) \\ in Struct \end{tabular}} & \multicolumn{1}{c}{Notes} \\
    \midrule
    \multirow{5}{*}{\smallnodedef{f : elem}} & \multirow{2}{*}{\smallcode{int32_t fc}} & Coordinate of nonzero (may be -1 if no nonzero stored) \\ \cmidrule{2-3}
    & \multirow{3}{*}{\smallcode{S fv}} & \smallcode{S} is type of nonzero value or pointer to data structure storing nonempty subtensor \\
    \smallnodedef{f : node_type} & \smallcode{node_type* f} & \\
    \smallnodedef{f : size} & \smallcode{int32_t f} \\
    \multirow{2}{*}{\smallnodedef{f : parent}} & \multirow{2}{*}{\smallcode{T* f}} & \smallcode{T} is struct type being emitted (or supertype, if applicable) \\
    \smallnodedef{f : bool} & \smallcode{bool f} \\
    \smallnodedef{f : [u]intN} & \smallcode{[u]intN_t f} & \smallcode{N} $\in \{8, 16, 32, 64\}$ \\
    \bottomrule
  \end{tabularx}
  }
  \label{tab:node-struct-fields}
\end{table}
}

\subsection{Generating Map Functions}
\label{sec:generating-maps}

When the nonzero coordinates of a tensor algebra computation's result is known to be a subset of an operand $T$'s nonzero coordinates, the computation can always be performed by simply mapping over and computing with each of $T$'s nonzeros.
%$T$'s nonzeros and computing with each nonzero.
%nonzeros in the result of a tensor algebra computation share the exact same coordinates as the nonzeros in a single dynamic sparse tensor operand, 
In particular, this is the case for all multiplicative operations, such as element-wise vector multiplication ($\iteration{i} a_i = b_i c_i$).
Thus, when one operand of such a computation is stored in a dynamic data structure while the rest are stored in formats that support efficient random access of nonzeros, our technique generates code that maps over the dynamic data structure in order to perform the computation.

To generate sequential code that can map over a dynamic data structure and perform some specific computation on the stored nonzeros, our technique emits a map function for every type of node in in the data structure.
\figref{bst-map-sequential} shows an example of a map function that our technique generates for mapping over nodes in a binary search tree and computing element-wise vector multiplication.
Each emitted function iterates over all \code{elem} fields in a node and, for each stored nonzero, perform the specified computation with the nonzero (as with lines 4--5 in \figref{bst-map-sequential}).
In the general case, the emitted function also iterates over all references to child nodes and, for each child node, (recursively) invokes the appropriate map function to process the node and its descendants (as with lines 6--7 in \figref{bst-map-sequential}).
By default, in order to better exploit cache spatial locality, our technique emits code that computes on all stored nonzeros in a node first before processing the node's descendants, as shown in \figref{bst-map-sequential}.
However, this optimization is not valid in situations where elements of a result tensor must be computed in coordinate order, such as if the output data structure requires elements to be appended in sequence.
In such cases, by using a node schema's sequence attribute as a guide, our technique can instead emit code to access (and compute on) a node's stored nonzeros and descendants in coordinate order.

\begin{figure}
  \centering
  \begin{minipage}[t]{0.49\columnwidth}
    \centering
    %\begin{tabular}{c}
      \begin{lstlisting}[basicstyle=\fontsize{8}{8}\ttfamily]
void map_b(bst* b, 
    double* a, double* c) {
  if (b) {
    int32_t i = b->ec;
    a[i] = b->ev * c[i];
    map_b(b->l, a, c);
    map_b(b->r, a, c);
  }
}


      \end{lstlisting}
    %\end{tabular}
    \vspace{-12pt}
    \subcaption {
      \label{fig:bst-map-sequential}
      Sequential BST map
    }
    \vspace{8pt}
  \end{minipage}
  \hfill
  \begin{minipage}[t]{0.49\columnwidth}
    \centering
    %\begin{tabular}{c}
      \begin{lstlisting}[basicstyle=\fontsize{8}{8}\ttfamily]
void map_b(blist* b, 
    double* a, double* c) {
  while (b) {
    for (int32_t p = 0; 
         p < b->B; p++) {
      int32_t i = b->ec[p];
      a[i] = b->ev[p] * c[i];
    }
    b = b->n;
  }
}
      \end{lstlisting}
    %\end{tabular}
    \vspace{-12pt}
    \subcaption {
      \label{fig:blist-map-sequential}
      Sequential block linked list
    }
    \vspace{8pt}
  \end{minipage}
    \begin{minipage}[t]{0.49\columnwidth}
    \centering
    %\begin{tabular}{c}
      \begin{lstlisting}[basicstyle=\fontsize{8}{8}\ttfamily]
void map_b(bst* b, 
    double* a, double* c,
    uint8_t d) {
  if (b) {
    if (l != 0) {
      #pragma omp task
      map_b(b->l, a, c, d-1);
      #pragma omp task
      map_b(b->r, a, c, d-1);
      int32_t i = b->ec;
      a[i] = b->ev * c[i];
    } else {
      map_b(b, a, c);
    }
  }
}
      \end{lstlisting}
    %\end{tabular}
    \vspace{-12pt}
    \subcaption {
      \label{fig:bst-map-parallel}
      Parallel BST map
    }
  \end{minipage}
  \hfill
  \begin{minipage}[t]{0.49\columnwidth}
    \centering
    %\begin{tabular}{c}
      \begin{lstlisting}[basicstyle=\fontsize{8}{8}\ttfamily]
void map_b(blist* b, 
    double* a, double* c) {
  while (b) {
    #pragma omp task
    for (int32_t p = 0; 
         p < b->B; p++) {
      int32_t i = b->ec[p];
      a[i] = b->ev[p] * c[i];
    }
    b = b->n;
  }
}




      \end{lstlisting}
    %\end{tabular}
    \vspace{-12pt}
    \subcaption {
      \label{fig:blist-map-parallel}
      Parallel block linked list map
    }
  \end{minipage}
  \caption {
    Examples of map functions that our technique emits.
    Note that Cilk-parallelized code can be similarly generated by replacing OpenMP pragmas with Cilk keywords.
  }
  \label{fig:map-function-examples}
\end{figure}

The above approach, in principle, generates correct code for any dynamic data structure that can be expressed using the node schema language.
That said, for data structures like linked lists that do not exhibit any fanout (i.e., those comprised of nodes that each only have one child), this approach can cause the stack to overflow at run-time if the input contains too many nodes.
Thus, for any type of node that (from statically analyzing its schema) is known to have exactly one child of the same type, our technique conceptually applies tail call optimization to instead emit a map function that uses a loop to iterate over all of the input node's descendants.
\figref{blist-map-sequential} shows an example of code that our technique emits for mapping over a block linked list using this approach.

Our technique also generates parallelized map functions in a similar way as sequential map functions.
For any node that has exactly one child of the same type (e.g., block linked list nodes), our technique parallelizes the processing of its descendants by emitting code that spawns a new task to compute on nonzeros stored in each node (as with lines 4--9 in \figref{blist-map-parallel}).
Meanwhile, for all other types of nodes, our technique emits code that spawns new parallel tasks to map over each child node and its descendants (as with lines 6--9 in \figref{bst-map-parallel}).
To avoid spawning too many fine-grained tasks, the emitted code keeps track of the depth of recursion (parameter \code{d} in \figref{bst-map-parallel}) and, once a certain depth has been reached, switches back to a sequential version of the map function (as with lines 12--13 in \figref{bst-map-parallel}).

Finally, to support mapping over nodes that are subtypes of some supertype, our technique emits a map function for each supertype that simply checks the input node's concrete type and invokes the concrete type's map function in order to actually compute on the input node.
So to map over a child of a B-tree node, for instance, the generated code would invoke a map function that takes any instance of \code{btree} as argument.
This function would, in turn, simply invoke a second map function (which performs the actual computation) that only takes an instance of either \code{btree_internal} or \code{btree_leaf} as argument, depending on if the child is an internal node (i.e., if \code{tp == btree_internal}) or a leaf node.

\subsection{Generating Iterators}
\label{sec:generating-iterators}

In general though, performing a tensor algebra computation may require simultaneously iterating over multiple operands that are all stored in dynamic data structures, which cannot be reasonably done with map functions.
To support such computations, our technique emits code that uses a set of loops to iterate over intersections or unions of the operands' nonzeros and compute with those nonzeros, as \figref{bst-elwise-mul-example} demonstrates for instance.
\citet{kjolstad2017} and \citet{henry2021} describe how such loops can be generated assuming it is possible to enumerate the stored nonzeros of each operand.
However, while \citet{chou2018} show how code to perform such enumeration can be emitted for operands that are stored in static, array-based formats, their technique does not support dynamic, pointer-based sparse tensor formats.

To generate an iterator that can enumerate the stored nonzeros in a dynamic data structure, our technique first mechanically emits a (recursive) coroutine for every type of node that may be contained in the data structure.
\figref{iterator-unoptimized} shows an example of such a coroutine, which our technique emits for iterating over a BST.
Each emitted coroutine accesses all of the input node's stored nonzeros and child nodes in the order specified by the node's sequence attribute.
For each nonzero, the emitted code simply yields the coordinate and value of that nonzero (as with line 5 in \figref{iterator-unoptimized}).
For each child node, the emitted code (recursively) invokes the appropriate coroutine to yield all nonzeros stored in the child node and its descendants (as with lines 4 and 6 in \figref{iterator-unoptimized}).

\begin{figure}
  \centering
  \begin{minipage}[t]{0.48\columnwidth}
    \centering
    %\begin{tabular}{c}
    \begin{minipage}[t]{\linewidth}
      \centering
      \begin{lstlisting}[basicstyle=\fontsize{8}{8}\ttfamily]
pair<int32_t,double>
iter_bst(bst* n) {
  if (n) {
    yield iter_bst(n->l);
    yield {n->ec, n->ev};
    yield iter_bst(n->r);
  }
}
      \end{lstlisting}
      \vspace{-12pt}
      \subcaption {
        \label{fig:iterator-unoptimized}
        Unoptimized iterator
      }
      \vspace{12pt}
    \end{minipage}
    \begin{minipage}[t]{\linewidth}
      \centering
      \begin{lstlisting}[basicstyle=\fontsize{8}{8}\ttfamily]
      
pair<int32_t,double>
iter_bst(bst* n) {
  while (n) {
    if (n->l)
      yield iter_bst(n->l);
    yield {n->ec, n->ev};
    n = n->r;
  }
}
      \end{lstlisting}
      \vspace{-12pt}
      \subcaption {
        \label{fig:iterator-optimized}
        After tail call optimization and null guard insertion
      }
    \end{minipage}
    %\end{tabular}
  \end{minipage}
  \hfill
  \begin{minipage}[t]{0.48\columnwidth}
    \centering
    %\begin{tabular}{c}
    \begin{lstlisting}[basicstyle=\fontsize{8}{8}\ttfamily]
pair<int32_t,double>
iter_bst(bst* n) {
  call_stack<uint8_t,bst*> s;
  s.emplace(0, n);
  while (!s.empty()) {
    n = get<1>(s.top());
    if (get<0>(s.top()) == 1)
      goto call_resume1;
    while (n) {
      if (n->l) {
        get<0>(s.top()) = 1;
        get<1>(s.top()) = n;
        s.emplace(0, n->l);
        goto call_end;
call_resume1:;
      }
      yield {n->ec, n->ev};
      n = n->r;
    }
    s.pop();
call_end:;
  }
}
    \end{lstlisting}
    %\end{tabular}
    \vspace{-12pt}
    \subcaption {
      \label{fig:iterator-norecurse}
      After recursion elimination
    }
  \end{minipage}
  \caption {
    Steps involved in generating an optimized iterator for BSTs.
    The final code is shown in \figref{bst-elwise-mul-example}.
  }
  \label{fig:iterator-generation}
\end{figure}

Our technique then applies a set of optimizations to each emitted coroutine in order to obtain a significantly more optimized iterator.
First our technique applies tail call optimization and inserts null checks around accesses to child nodes in order to reduce the number of recursive calls.
Additionally, if the input node has child nodes of other types, all invocations of iterators for those nodes are inlined, yielding a coroutine that only has recursive calls to itself.
Then, to eliminate the overhead of recursive calls to a coroutine, our technique rewrites the coroutine so that it emulates recursion using a loop with an explicit call stack, which stores the local variables and state of each recursive call.
Finally, to obtain code that does not rely on language support for coroutines (and that can thus be compiled with pre-C++20 compilers or even trivially translated to C), our technique rewrites the coroutine to a function that, when invoked, yields the next nonzero's coordinate and value as output parameters.
\figref{iterator-generation} show how our technique applies these optimizations to the unoptimized code in \figref{iterator-unoptimized} in order to generate an efficient iterator for BSTs, which is shown in \figref{bst-elwise-mul-example}.
Then, by applying the technique of \citet{kjolstad2017}, our technique can emit code that invokes this iterator to iterate over a BST simultaneously with any other static or dynamic data structure, including another BST (as in \figref{bst-elwise-mul-example}) or an array-based sparse vector (as in \figref{append-unspecialized-example}).

\subsection{Generating Assembly Code}
\label{sec:generating-assembly}

In addition to generating code that compute on sparse tensor operands stored in dynamic tensor formats, our technique can emit code to store the results of computations in dynamic tensor formats as well.
This is achieved in several ways.

\setlength{\intextsep}{5pt}%
\setlength{\columnsep}{10pt}%
\begin{wrapfigure}{R}{0.48\columnwidth}
\begin{lstlisting}[basicstyle=\fontsize{8}{8}\ttfamily]
bst* map_b(bst* b, 
    double* c) {
  if (b) {
    bst* ret = new bst;
    int32_t i = b->ec;
    ret->ec = i;
    ret->ev = b->ev * c[i];
    ret->l = map_b(b->l, c);
    ret->r = map_b(b->r, c);
    return ret;
  }
  return NULL;
}
\end{lstlisting}
    \caption{
        Example generated map function that assembles a dynamic data structure to store the result.
    }
    \label{fig:map-with-assembly}
\end{wrapfigure}

If the computation can be performed with a map function (as described in \secref{generating-maps}), and if the result is also stored in the same format as the input tensor being mapped over, then our technique emits a map function that assembles the result by essentially deep copying the input data structure.
This approach is valid since each nonzero in the result is computed from one nonzero in the input tensor being mapped over, so our technique can infer that the output data structure must have the same structure as the input data structure.
\figref{map-with-assembly} shows an example map function that our technique generates, which computes on an input tensor that is stored as a BST and which stores the result as another BST.
Such map functions can be generated in largely the same way as described in \secref{generating-maps}.
To deep copy the input data structure though, each emitted map function additionally allocates and returns a new node that is of the same type as the input node (lines 4 and 10 in \figref{map-with-assembly}).
This new node is initialized by copying over the coordinates of each input nonzero (line 6 in \figref{map-with-assembly}), with the corresponding values initialized to be the results of the computation (line 7 in \figref{map-with-assembly}).
Furthermore, new child nodes are allocated for the new output node by invoking the augmented map function(s) on the input node's children (lines 8--9 in \figref{map-with-assembly}).

In general though, a tensor algebra kernel may have to assemble a dynamic data structure from scratch to store the result.
By using the abstraction we propose in \secref{assembly-interface}, our technique is able to generate such code without needing to hard-code for any specific data structure.
Specifically, to generate code that stores the result of a computation in a dynamic data structure, our technique first emits code that invokes the assembly functions described in \secref{assembly-interface} to store the result nonzeros.
Then, the emitted code is specialized to a specific type of dynamic data structure by inlining its implementation of the assembly functions.
So to generate code that stores the result of a tensor algebra computation in a block linked list, for instance, our technique emits code like what is shown in \figref{append-unspecialized-example}, which invokes the \code{append_first} and \code{append_rest} functions to store the result nonzeros.
The code generator can then inline implementations of \code{append_first} and \code{append_rest} for block linked lists (as shown in \figref{append-example}) into the emitted code, yielding code that is specialized for block linked list outputs.
In the same way, if a computation simply assigns an input tensor to the output and if the input is stored in an array-based format, our technique can emit code that invokes the \code{build} function (with a reference to the input as the argument \code{elems}) to bulk assemble the output.
The code generator can then inline any dynamic tensor format's implementation of \code{build} in order to obtain code that bulk assembles the output in that format.

\begin{figure}
\begin{lstlisting}[basicstyle=\fontsize{8}{8}\ttfamily]
  blist_head* aret;
  bool afirst = true;
  uint8_t bstate = iter_bst(0, ..., ib, bval);
  int32_t pc = c_pos[0];
  while (bstate != 0 && pc < c_pos[1]) {
    int32_t ic = c_crd[pc];
    int32_t i = min(ib, ic);
    if (i == ib && i == ic) {
      double aval = bval * c_vals[pc];
      if (afirst) {
        aret = new blist_head;
        append_first({i, aval}, astate, aret);
        afirst = false;
      } else {
        append_rest({i, aval}, astate);
      }
    }
    if (i == ib)
      bstate = iter_bst(bstate, ..., ib, bval);
    pc += (i == ic);
  }
\end{lstlisting}
  \caption{
    Example emitted code that invokes \code{append_first} and \code{append_rest} to store result nonzeros.
    Our technique can specialize this code for a specific output format by inlining that format's implementations of the append functions.
  }
  \label{fig:append-unspecialized-example}
\end{figure}

\section{Evaluation}
\label{sec:evaluation}
We implement our technique as a prototype extension to the TACO sparse tensor algebra compiler and find it generates efficient sparse tensor algebra kernels that compute on operands stored in dynamic tensor formats.
Code that our technique emits have performance comparable to or better than equivalent code that are either directly implemented in hand-optimized libraries or implemented using hand-optimized primitives provided by libraries.
Furthermore, we find that our technique emits efficient code to simultaneously compute on dynamic and static sparse tensors, which enables our technique to outperform hand-implemented libraries that only support operations on dynamic data structures.

\subsection{Experiment Setup}
\label{sec:experiment-setup}

We evaluate code that our technique generates against hand-optimized code that are implemented in Aspen~\cite{aspen} and PAM~\cite{pam}.
Aspen is a state-of-the-art C++ graph processing framework, which allows users to implement applications that compute on dynamic graphs (stored as C-trees) by invoking a fixed set of primitives for mapping over and applying user-defined functions on edges and vertices.
PAM, on the other hand, is a lower-level parallel C++ library that supports a fixed set of primitives for operating on ordered key-value maps stored as self-balancing BSTs.
While PAM does not directly implement any (multidimensional) tensor algebra kernel, PAM exposes primitives that can be used to compute tensor algebra operations on operands that are stored using BSTs; indeed, Aspen is also implemented on top of PAM.

We run our experiments on a two-socket, 12-core/24-thread 2.5 GHz Intel Xeon E5-2680 v3 machine with 30 MB of L3 cache per socket and 128 GB of main memory. 
The machine runs Ubuntu 18.04.3 LTS.
We compile all code using GCC 7.5.0 with \code{-O3 -march=native} optimizations enabled and use Cilk for parallel execution.
To ensure an apples-to-apples comparison of the actual algorithms that the (generated and hand-optimized) tensor algebra kernels implement, we modify code generated by our technique so that they operate on the exact same data structures in memory as Aspen and PAM;.
(This only requires minor changes to how the fields of input nodes are accessed and does not entail any algorithmic change.)
Additionally, all memory allocations are done using \code{jemalloc}.
We run each experiment 100 times under cold cache conditions and report median execution times.
Each experiment is run using 24 threads, with execution restricted to a single socket using \code{numactl}.

We run our experiments with real-world sparse matrices from the SuiteSparse Matrix Collection~\cite{suitesparse}.
These matrices, which \tabref{input-summary} describes in more detail, represent graphs and other data that arise in disparate application domains.

{
\setlength{\tabcolsep}{4.9pt}
\begin{table}
  \caption{
    Statistics about matrices used in our experiments.
  }
  \centering
  {\small
  \begin{tabular}{lllr}
    \toprule
    & \multicolumn{1}{c}{Matrix} & \multicolumn{1}{c}{Dimensions} & \multicolumn{1}{c}{NNZ} \\
    \midrule
    1 & belgium\_osm & 1.44M $\times$ 1.44M & 3.10M \\
    2 & cit-Patents & 3.77M $\times$ 3.77M & 16.5M \\
    3 & coAuthorsCiteseer & 227K $\times$ 227K & 1.63M \\
    4 & com-Orkut & 3.07M $\times$ 3.07M & 234M \\
    5 & coPapersDBLP & 540K $\times$ 540K & 30.5M \\
    6 & delaunay_n24 & 16.8M $\times$ 16.8M & 101M \\
    7 & indochina-2004 & 7.41M $\times$ 7.41M & 194M \\
    8 & rgg_n_2_24_s0 & 16.8M $\times$ 16.8M & 265M \\
    9 & road_central & 14.1M $\times$ 14.1M & 33.9M \\
    10 & road_usa & 23.9M $\times$ 23.9M & 57.7M \\
    11 & roadNet-CA & 1.97M $\times$ 1.97M & 5.53M \\
    12 & ship_003 & 122K $\times$ 122K & 3.78M \\
    13 & soc-LiveJournal1 & 4.85M $\times$ 4.85M & 69.0M \\
    14 & webbase-1M & 1.00M $\times$ 1.00M & 3.11M \\
    \bottomrule
  \end{tabular}
  }
  \label{tab:input-summary}
\end{table}
}

\subsection{Performance Evaluation}
\label{sec:performance-evaluation}

We first evaluate code that our technique emits for computing the main kernel in each iteration of the PageRank algorithm~\cite{Pageetal98}, which can be expressed in concrete index notation as $\titeration{i}\iteration{j} y_i \reduce{+} A_{ij}x_j{d_j}^{-1}$, where $A$ represents a graph's adjacency matrix and $y$, $x$, and $d$ are dense vectors.
Specifically, we measure the performance of code that our technique emits for $A$ stored using only BSTs (i.e., in the (\code{bst}, \code{bst}) format) and for $A$ stored as C-trees (i.e., in the (\code{bst}, \code{ctree}) format).
We then compare the generated code against Aspen, which implements an \code{edgeMap} primitive that supports the same PageRank kernel for $A$ stored as C-trees.\footnote{While Aspen supports C-trees that use difference encoding to compress the coordinates stored in each block, we only evaluate our technique and Aspen on C-trees that do not use difference encoding, since difference encoding is not supported by our technique as we have described it.}
Additionally, we compare the generated code against PAM, which can compute the same kernel (for $A$ stored as BSTs) by performing a map operation (\code{map_void}) over the rows of $A$ that, in turn, performs a map-reduce operation over the nonzeros in each row of $A$ to compute an entry of $y$.

We also evaluate code that our technique emits for computing two additional sparse tensor algebra kernels on a dynamic matrix $A$ that is stored as BSTs:
\begin{itemize}
  \item Sparse matrix-vector multiplication ($\titeration{i}\iteration{j} y_i \reduce{+} A_{ij}x_j$) with the result $y$ also stored as a BST
  \item Sparse matrix addition ($\titeration{i}\iteration{j} C_{ij} = A_{ij} + B_{ij}$) with $B$ and $C$ being static matrices stored in CSR
\end{itemize}
We do not compare against Aspen as it does not support these kernels (on matrices stored as C-trees or BSTs).
However, we compare the generated code against PAM, which implements primitives that can be utilized to compute both kernels.
In particular, PAM can be used to compute sparse matrix-vector multiplication in a similar way as the PageRank kernel, except that the map operation (\code{map}) over $A$'s rows also constructs a new BST to store the result values.\footnote{While PAM uses a custom pool allocator to allocate new BST nodes by default, we modify PAM for our experiments so that it simply uses \smallcode{malloc} to allocate new nodes. We find that, for our benchmarks, this slightly improves PAM's performance and also yields more repeatable performance results.}
Meanwhile, sparse matrix addition can be computed row by row by having PAM convert each row of $B$ to a BST, compute the union of the BST with the corresponding row in $A$ (using \code{map_union}), and map over the result (which is also in a BST) to copy each result nonzero to $C$ (using \code{foreach_index}).

\begin{figure}
  \centering
  \includegraphics[width=0.99\columnwidth]{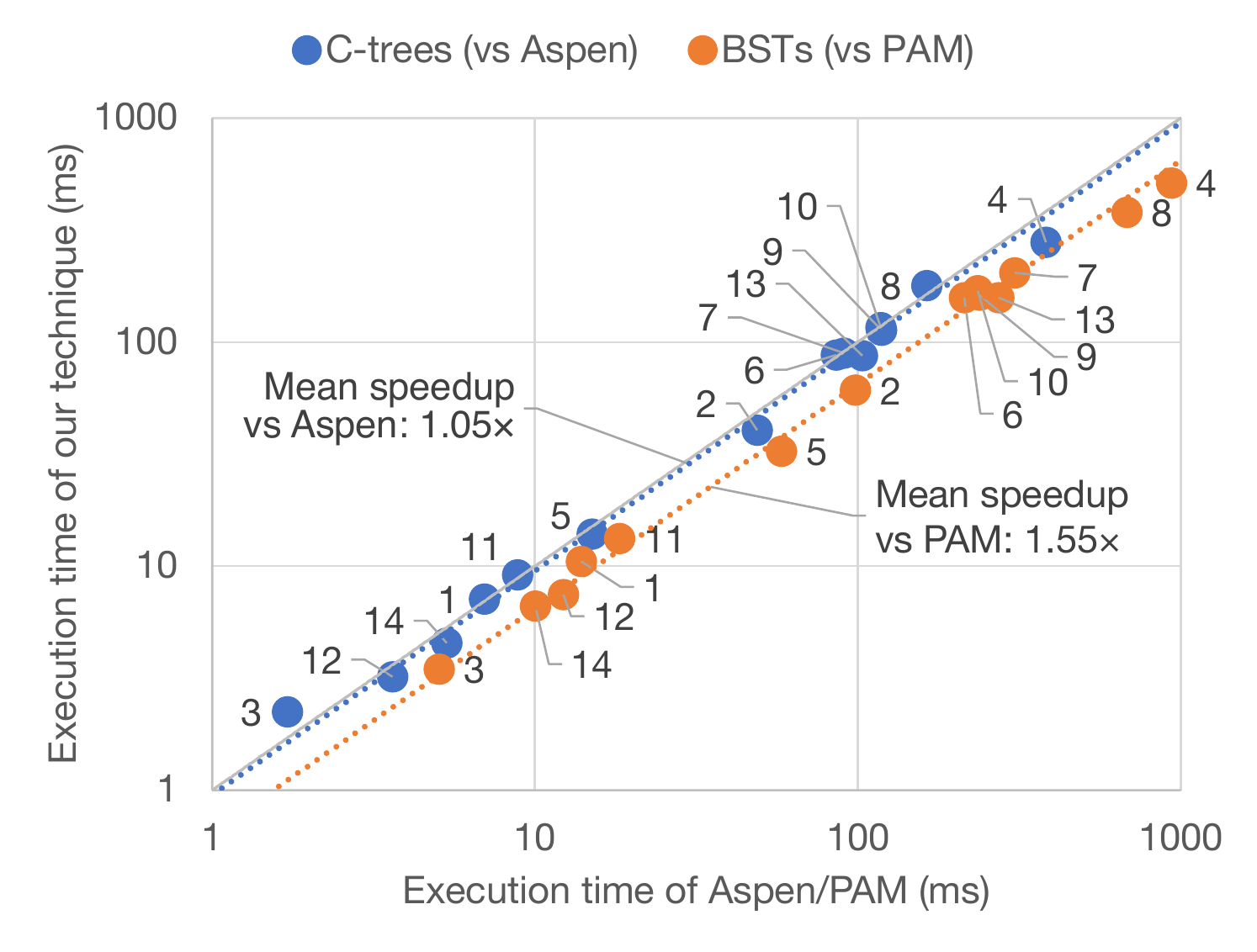}
  \caption {
    Performance of PageRank kernels that are generated by our technique or implemented using Aspen or PAM, with input matrices stored as BSTs (supported by our technique and PAM) or as C-trees (supported by our technique and Aspen).
    Each data point represents an experiment with an input matrix listed in \tabref{input-summary}.
    Data points that lie below the diagonal line gray indicate that code generated by our technique runs faster than Aspen or PAM.
  }
   \label{fig:pagerank-results}
\end{figure}

\begin{figure}
  \centering
  \includegraphics[width=0.99\columnwidth]{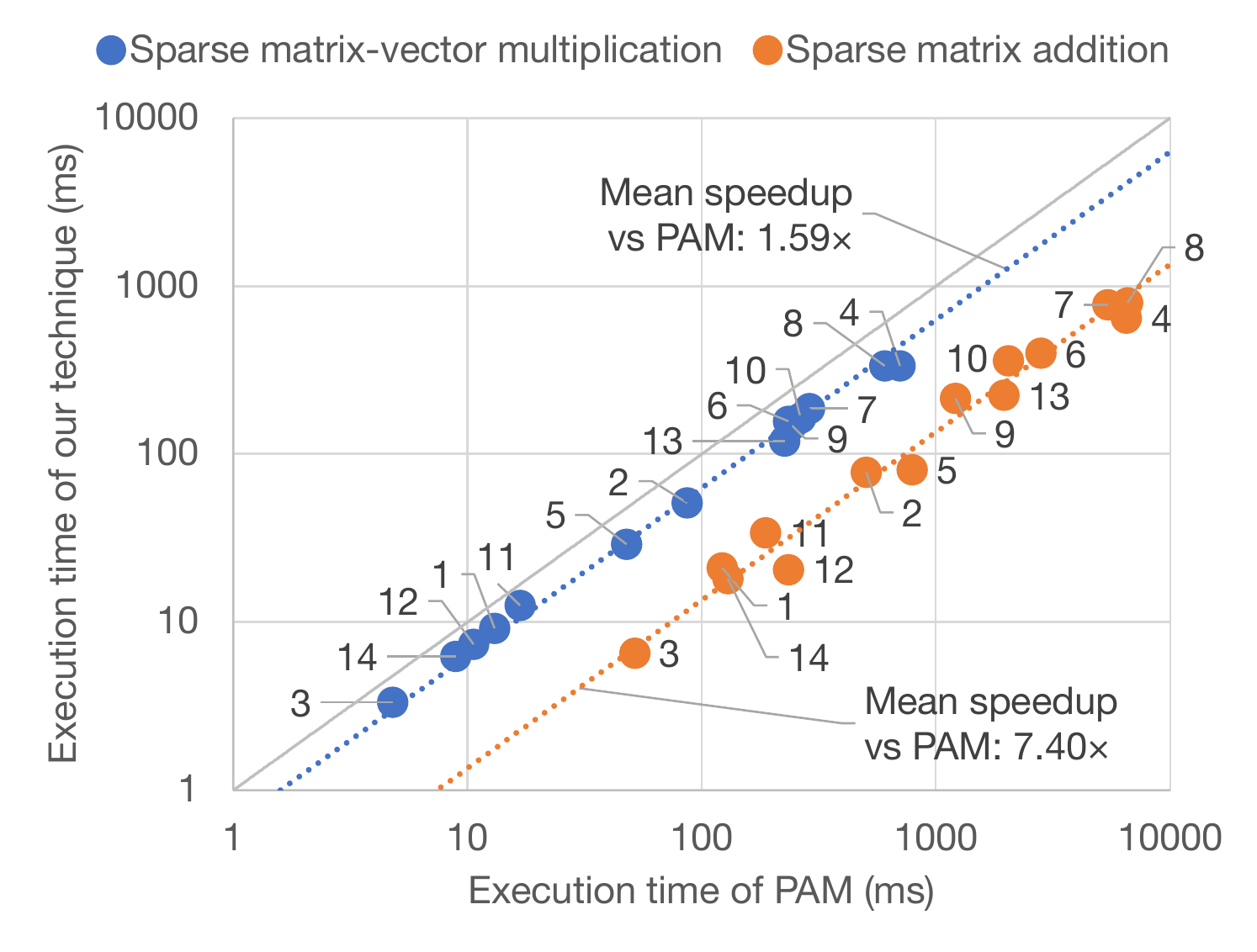}
  \caption {
    Performance of sparse matrix-vector multiplication and sparse matrix addition kernels that are generated by our technique or implemented using PAM.
    Each data point represents an experiment with a matrix listed in \tabref{input-summary}.
    Data points that lie below the gray diagonal line indicate that code generated by our technique runs faster than PAM.
  }
   \label{fig:misc-results}
\end{figure}

\figsref{pagerank-results}{misc-results} show the results of our experiments, which demonstrate that our technique generates efficient dynamic sparse tensor algebra kernels.
On the whole, code that our technique generates has similar performance as Aspen, with the generated PageRank kernel being 1.05$\times$ faster than Aspen on average.
This is unsurprising since our technique emits code that implements the same high-level algorithm as Aspen's (dense) \code{edgeMap} primitive (though with a slightly different approach for achieving load-balancing, which accounts for the variations in relative performance with individual matrices).
On the other hand, for both the PageRank kernel and sparse-matrix vector multiplication with a BST result, our technique generates code that outperforms PAM by 1.55$\times$ and 1.59$\times$ on average.
PAM incurs additional overhead for these kernels since its implementation of map-reduce always performs two addition operations for every node in a tree, even if a node has fewer than two children (making some of those additions unnecessary).
By contrast, our technique generates more efficient code that only performs one addition operation for every node.

Additionally, for sparse matrix addition of a dynamic matrix with a static matrix, our technique emits code that outperforms equivalent code implemented with PAM by 7.40$\times$ on average.
Performing the computation using PAM incurs significant overhead since the library only supports computing unions of maps that are stored as BSTs.
Thus, in order to compute each row of the result, many additional memory operations are needed to allocate new nodes when converting a row of the static CSR input matrix to a BST and also when actually performing the union operation.
Furthermore, using PAM requires additional overhead in order to copy values computed by the union operation over to the actual output matrix, which is also stored in CSR.
By contrast, our technique emits code that merges the two input matrices by simultaneously iterating over their nonzeros and that directly stores the result nonzeros into the CSR output without needing a BST temporary.
This shows the benefits of a compiler technique like ours that can generate efficient code to compute with both static and dynamic sparse tensors.

\subsection{Comparison with Static Sparse Tensor Formats}
\label{sec:static-format-comparison}

\begin{figure}
  \centering
  \includegraphics[width=0.99\columnwidth]{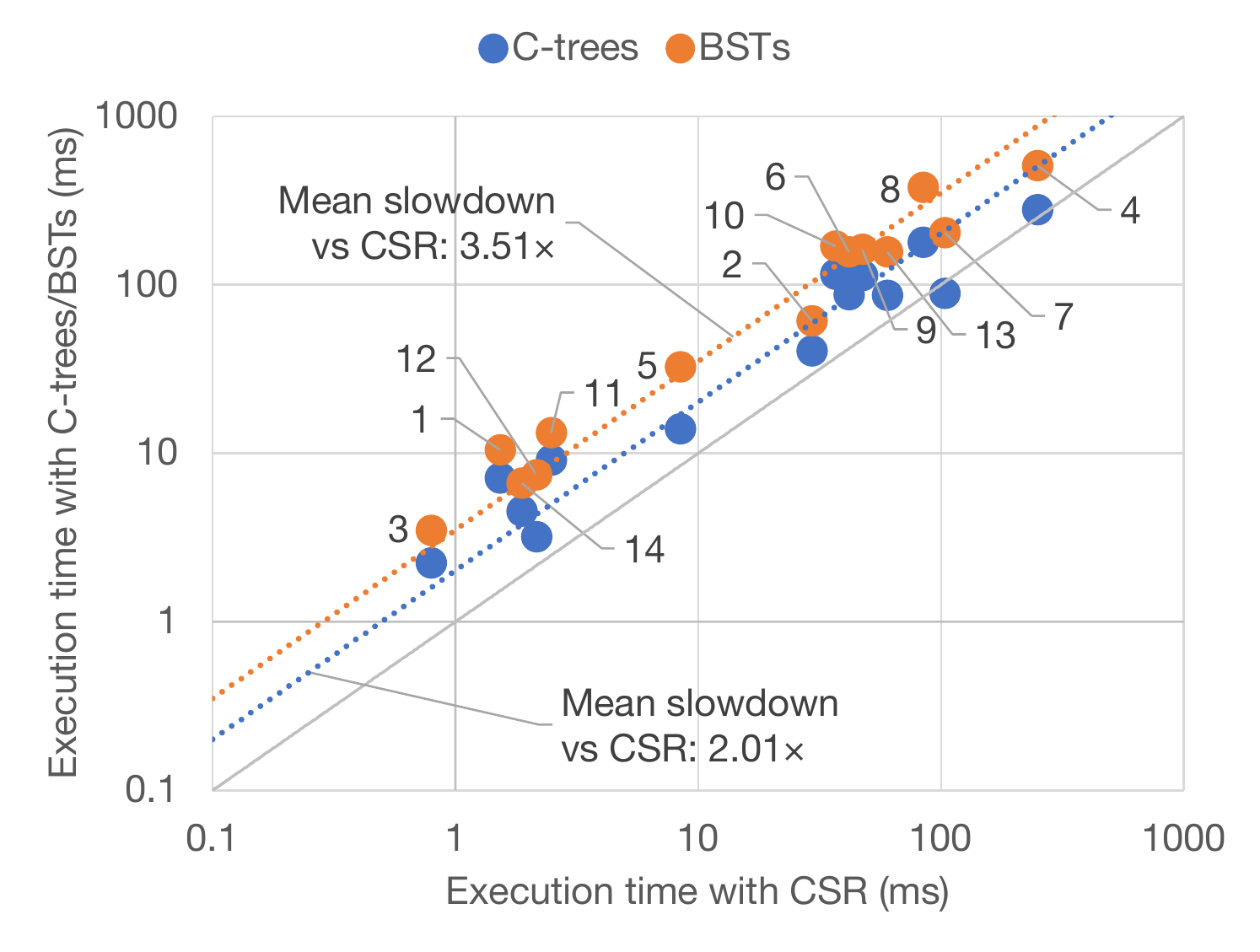}
  \caption {
    Performance of generated PageRank kernels that compute on input matrices stored in dynamic data structures (C-trees or BSTs) and a static matrix format (CSR).
    Each data point represents an experiment with a matrix listed in \tabref{input-summary}.
    Data points that lie above the gray diagonal line indicate that computing with C-trees or BSTs is slower than with CSR.
  }
   \label{fig:static-perf-comparison}
\end{figure}

We further compare code that our technique emits for computing on dynamic sparse tensors against code that TACO (without our extension) emits for computing on static sparse tensors.
As \figref{static-perf-comparison} shows, code that our technique emits for computing the PageRank kernel on matrices stored as C-trees and BSTs are only 2.01$\times$ and 3.51$\times$ slower on average, respectively, than code that computes the same kernel on CSR matrices.
On the other hand though, as alluded to in \secref{introduction}, inserting a nonzero into a CSR matrix can be as expensive as actually computing with the matrix, whereas inserting into BSTs or C-trees is asymptotically much cheaper.
The results in \figref{static-perf-comparison} thus imply that, for applications that need to repeatedly compute on dynamic sparse tensors, the overhead of storing the tensors in dynamic data structures can be easily amortized.

\section{Related Works}
\label{sec:related-works}
As \secref{dynamic-formats-background} illustrates, there exists a long line of works on using dynamic data structures to efficiently represent adjacency matrices of dynamic graphs.
In addition to the libraries and frameworks summarized in \secref{dynamic-formats-background}, LLAMA~\cite{llama} uses a data structure that resembles variable block linked lists to store entries of a dynamic graph's adjacency matrix, except that each block can store entries corresponding to multiple rows.
There are also various GPU libraries and frameworks that store dynamic graphs using variants of either the data structures shown in \figref{dynamic-tensor-formats}~\cite{aimgraph,slabhash} or other data structures that can be expressed using our proposed abstractions~\cite{hornet}.
Furthermore, a number of works~\cite{dynamiccsr,evograph} have explored using array-based data structures, including packed memory arrays~\cite{terrace,gpma}, to store dynamic graphs.
However, all these works, including those summarized in \secref{dynamic-formats-background}, rely on hand-optimized kernels to compute on graphs stored in their data structure of choice, whereas our technique instead automatically generates such kernels.

Existing sparse linear and tensor algebra compilers cannot readily, if at all, generate code to efficiently compute on tensors stored in disparate dynamic tensor formats.
TACO~\cite{kjolstad2017,kjolstad2019,chou2018,chou2020} emits efficient sparse tensor algebra kernels that compute with static sparse tensors stored in a wide range of array-based formats like CSR and DIA~\cite{saad2003}.
However, TACO cannot generate code to compute with dynamic sparse tensors that are stored in pointer-based formats, since the sparse tensor format abstraction of \citet{chou2018} cannot represent those data structures.
Bernoulli~\cite{kotlyar-thesis,stodghill-thesis,bernoulli} similarly generates sparse linear algebra kernels using an abstraction for sparse vector and matrix formats called the black-box protocol.
\citet{kotlyar-thesis} show how array-based linked lists can be expressed using the black-box protocol, though they do not consider other pointer-based data structures such as BSTs.
Furthermore, the black-box protocol requires a user to implement low-level iterators for data structures, whereas our technique can automatically generate optimized iterators for dynamic data structures.
MT1~\cite{bik1,bik2,Bik1996CompilerSF} and SIPR~\cite{sipr}, meanwhile, each only support a fixed set of array-based formats for storing sparse vectors and matrices and do not support any pointer-based data structures.
More recently, \citet{chill} have shown how polyhedral techniques can be utilized to generate sparse linear algebra code by representing array-based sparse matrix formats as uninterpreted functions.
Furthermore, \citeauthor{Arnold:2010:SVS:1863543.1863581}~\cite{ll,Arnold:2010:SVS:1863543.1863581} have shown how computations on array-based sparse matrix formats can be expressed using a functional language they develop called LL.
Again though, these techniques cannot generate the types of algorithms that are needed to compute with dynamic tensors stored in recursive, pointer-based data structures.

There exists a separate line of works on synthesizing data structure operations from declarative specifications.
Many techniques have been proposed for synthesizing imperative programs that modify pointer-based data structures like AVL trees and linked lists, given either user-specified invariants~\cite{kurilova2013,qiu2017} or graphical specifications of the desired programs' inputs and outputs~\cite{singh2011}.
Other techniques have also been proposed for synthesizing functional programs from declarative specifications, including programs that process and manipulate pointer-based data structures~\cite{kneuss2013,polikarpova2016}.
None of these techniques consider block data structures like C-trees, and they do not generate parallel code.
In addition, \citet{rayside2012} show how Java iterators can be synthesized for pointer-based data structures given specifications written in relational logic, though their technique does not generate map functions or any other code to actually compute with elements stored in these data structures.

\section{Conclusion}
\label{sec:conclusion}
We have shown how a compiler can automatically generate efficient code to perform tensor algebra computations on dynamic sparse tensors that are stored in recursive, pointer-based data structures.
Simply by specifying how these data structures organize nonzeros in memory and how these data structures can be assembled, a user can extend our compiler to support new dynamic tensor formats without having to modify the code generator itself. 
This makes it possible to efficiently compute on dynamic sparse tensors that are stored in disparate formats, which can make our technique more applicable to a wider range of applications and domains.

% Acknowledgments
\begin{acks}
  This work was supported by the Application Driving Architectures (ADA) Research Center, a JUMP Center co-sponsored by SRC and DARPA; the U.S. Department of Energy, Office of Science, Office of Advanced Scientific Computing Research under Award Numbers DE-SC0008923 and DE-SC0018121; and DARPA under Awards HR0011-18-3-0007 and HR0011-20-9-0017.
  Any opinions, findings, and conclusions or recommendations expressed in this material are those of the authors and do not necessarily reflect the views of the aforementioned funding agencies.
\end{acks}

\balance
\bibliography{paper} 

\end{document}